\documentclass{elsart}

\pdfoutput=1

\journal{Digital Signal Processing}

\usepackage{amssymb}
\usepackage{amsmath}

\newif\ifpdf
\ifx\pdfoutput\undefined
\pdffalse 
\else
\pdfoutput=1 
\pdftrue
\fi

\ifpdf
\usepackage[pdftex]{graphicx}
\usepackage{epstopdf}
\else
\usepackage{graphicx}
\fi

\ifpdf
\graphicspath{{./figures/}}
\DeclareGraphicsExtensions{.jpg,.pdf}
\else
\graphicspath{{./eps/}}
\DeclareGraphicsExtensions{.eps, .jpg}
\fi

\newcommand{\WV}{Wigner-Ville}
\newcommand{\stft}{short-time Fourier transform}
\newcommand{\stfts}{short-time Fourier transforms}
\newcommand{\tfr}{time-frequency representation}
\newcommand{\tfrs}{time-frequency representations}
\newcommand{\Tfrs}{Time-frequency representations}
\newcommand{\mwt}{moving window transform}
\newcommand{\fir}{finite impulse response}
\newcommand{\dft}{discrete Fourier transform}
\newcommand{\ft}{Fourier transform}
\newcommand{\rabwe} {re\-assigned bandwidth-enhanced}

\newcommand{\mmwm}{modified moving window method}

\newcommand{\picwidth}{5in}

\begin{document}

\begin{frontmatter}

\date{30 September 2005}



\title{A Unified Theory of Time-Frequency Reassignment}


\author[kel]{Kelly R. Fitz\corauthref{cor}},
\corauth[cor]{Corresponding author. Tel.: +1-952-947-4580}
\ead{kelly\_fitz@starkey.com}
%
\author[sean]{Sean A. Fulop}
\ead{sfulop@csufresno.edu}

\address[kel]{Starkey Labs \\
6600 Washington Ave South, Eden Prairie MN 55344}
\address[sean]{Department of Linguistics PB92, California State University, \\
5245 N. Backer Ave. Fresno, CA, 93740-8001.}

\begin{abstract}
\Tfrs\ such as the spectrogram are commonly used 
to analyze signals having a time-varying
distribution of spectral energy, but the spectrogram is
constrained by an unfortunate tradeoff between resolution in time and 
frequency. A method of achieving high-resolution 
spectral representations has been 
independently introduced by several parties. The technique has been
variously named {\em reassignment} and {\em remapping}, but while the
implementations have differed in details, they are all based on the
same theoretical and mathematical foundation. 
In this work, we present a brief history of work on the method 
we will call {\em the method of time-frequency 
reassignment}, and present a unified mathematical description
of the technique and its derivation. We will focus on the development
of time-frequency reassignment in the context of
the spectrogram, and conclude with a discussion
of some current applications of the reassigned spectrogram. 
%
\end{abstract}

\end{frontmatter}


\section{Introduction}

Many signals of interest
have a distribution of energy that varies in time and frequency. 
For example, any sound signal having a beginning or an end has an
energy distribution that varies in time, and
most sounds exhibit considerable variation in both
time and frequency over their duration. 
\emph{\Tfrs} are commonly used 
to analyze or characterize such signals.
They map the one-dimensional time-domain signal
into a two-dimensional function of time and frequency.
A \tfr\ describes the variation of spectral
energy distribution over time, much as a musical score describes the
variation of musical pitch over time. 

In audio signal analysis, the \emph{spectrogram} is the most commonly-used
\tfr, probably because it is well-understood, and immune 
to so-called \emph{cross-terms} that sometimes make other \tfrs\ 
difficult to interpret. But the windowing operation required in spectrogram 
computation introduces an unsavory tradeoff between time
resolution and frequency resolution, so spectrograms provide a
\tfr\ that is blurred in time, in frequency, or in both dimensions.


\emph{Time-frequency reassignment} is a technique for 
refocussing time-frequency data in a blurred representation
like the spectrogram by mapping the data to time-frequency
coordinates that are nearer to the true region of support of
the analyzed signal. 
This method has been presented in other publications.
In particular, the reader is referred
to the excellent tutorial in~\cite{flandrin03time}
and to~\cite{fulop05algorithms}.
Time-frequency reassignment is gaining popularity, 
but it is still common to
see new research conducted using the classical spectrogram 
that could benefit in efficiency or effectiveness from the 
enhancements afforded by reassignment. 

In this paper, we will consider the application of time-frequency 
reassignment only to spectrogram data, though its effectiveness
has also been demonstrated in the context of many other representations
(see for example the extensive discussion of reassigned
\tfrs\ and time-scale representations in~\cite{auger95improving}).
For completeness, 
and because different formulations are popular in different
research communities, we begin with a review of time-frequency
analysis using the spectrogram in Section~\ref{sec:spectrogram}
before introducing the theory of time-frequency reassignment 
in Section~\ref{sec:method}. 
In Section~\ref{sec:if} we discuss the notion of instantaneous
frequency and its estimation using frequency reassignment. 
In Section~\ref{sec:separability}, we introduce the condition 
called ``separability'', 
which is crucial to obtain a meaningful \tfr\ of a multicomponent
signal. 

Most common applications of the method of time-frequency reassignment are
discrete (sampled) in time and frequency, so we next turn our attention to
methods for efficiently estimating or computing
reassigned time and frequency coordinates in the discrete case.
Algorithms for implementing time-frequency reassignment have been
presented in~\cite{flandrin03time} and~\cite{fulop05algorithms},
but complete derivations of the techniques have not been published, to our
knowledge, so in Section~\ref{sec:efficient},
we offer them as a useful starting point for future research 
in the computation of higher-order spectral derivatives. 

We then consider some applications of the method of reassignment. 
In Section~\ref{sec:pruning}, we discuss improvements in spectrogram
readability that can be achieved using reassigned data, with specific
attention to methods of detecting and removing noisy or unreliable 
spectral data. 
In sound modeling applications, not only the
spectral energy or magnitude is needed, but also the spectral phase.
In Section~\ref{sec:phase}, we discuss a high-fidelity additive
sound model that is constructed from reassigned spectral data, and
methods by which the \stft\ phase can be corrected to agree
with the with the time-frequency estimates obtained
by reassignment.
Finally, in Section~\ref{sec:higher}, we suggest some 
directions for future research using higher-order spectral 
derivatives.

\section{The Spectrogram as a Time-Frequency Representation}
\label{sec:spectrogram}

One of the best-known \tfrs\ is the {\em spectrogram},  
defined as the squared magnitude of the \stft
\begin{equation}
\label{eqn:specgram}
S(t,\omega) = |X(t,\omega)|^2.
\end{equation}
The \stft\ is defined as a complex function of continuous time $t$ and 
radian frequency $\omega$ by
\begin{align}
X(t,\omega) & = \int x(\tau) h^\ast(t-\tau)e^{-j\omega \tau}d\tau \\
\label{eqn:stft}
 & = \int x(\tau) h(t-\tau)e^{-j\omega \tau}d\tau \\
\label{eqn:Mphi} 
		& = M(t,\omega) e^{j \phi(t,\omega)}
\end{align}
where $h(t)$ is a finite-length, real-valued window function, 
(so $h(t)=h^\ast (t)$), 
$M(t,\omega)$ is the magnitude of
the short-time Fourier transform, and $\phi(t,\omega)$ is its phase.

Often, it is more convenient to compute the time-varying spectrum
by shifting the input signal, $x(t)$, instead of the window function.
This modified transform, computed by
\begin{align}
\label{eqn:mwt} 
X_{t}(\omega) & = \int x(\tau + t) h(-\tau)e^{-j\omega \tau}d\tau \\
		& = M_{t}(\omega) e^{j \phi_{t}(\omega)}
\end{align}
is simply the Fourier transform of the shifted and windowed 
input signal, $M_{t}(\omega)$ is the magnitude of
the Fourier transform, and $\phi_{t}(\omega)$ is its phase.

Equations~\ref{eqn:stft} and~\ref{eqn:mwt} differ
in the range of $\tau$, the variable of integration, 
over which the integrand is non-zero. Since
$h(t)$ is a finite-duration window function, the integrand 
in Equation~\ref{eqn:mwt} is always non-zero over the same
range of~$\tau$, for any $t$.
Thus, the temporal reference of the transform ``slides along''
the signal, instead of remaining fixed at $t=0$, so we can call this 
transform the {\em moving window transform}~\cite{kodera78analysis},
to distinguish it from the \stft. 


The fixed range of integration in Equation~\ref{eqn:mwt} makes the \mwt\
easy to implement directly in a digital system using a fast Fourier transform, 
but the two transforms are, in fact, equivalent, differing only in their temporal
reference.
By a change of variable, $t^{\prime} = \tau + t$, in Equation~\ref{eqn:mwt},
we can show that 
\begin{align}
X_{t}(\omega) & = \int x(\tau + t) h(-\tau)e^{-j\omega \tau}d\tau \\
		& = \int x(t^{\prime}) h(t - t^{\prime})e^{j\omega (t - t^{\prime})}dt^{\prime} \\
		& = e^{j\omega t}  \int x(t^{\prime}) h(t - t^{\prime}) e^{-j\omega t^{\prime}}dt^{\prime}\\
\label{eqn:stftrelate}
		& = e^{j\omega t} X(t,\omega)  \\
		&= M(t,\omega) e^{j \left[ \omega t + \phi(t,\omega) \right]}
\end{align}
so the \mwt\ defined by Equation~\ref{eqn:mwt} is closely related to
the \stft.
The magnitudes of the two transforms are
equal, and the phases differ only by a linear frequency term, that is,
\begin{align}
\label{eqn:Mequal}
M_{t}(\omega) &=  M(t,\omega)  \\
\phi_{t}(\omega) &= \omega t + \phi(t,\omega) 
\end{align}

For a sinusoid having constant frequency, $\omega_{0}$,
the phase of the \stft\ evaluated at that frequency,
$\phi(t,\omega_{0})$, is constant for all time, and equal to the
phase of the sinusoid at $t=0$, whereas $\phi_{t}(\omega_{0})$, 
the phase of the \mwt\ evalutated at frequency $\omega_{0}$, 
rotates at exactly the frequency of the 
sinusoid.

Though the short-time phase spectrum is known to contain important
temporal information about the signal, this information is difficult to interpret, so
typically, only the short-time magnitude spectrum
is considered in the construction of a \tfr\ like the spectrogram.
In the construction of additive sinusoidal sound models, the short-time phase
spectrum is sometimes used to improve the frequency estimates in the
\tfr\ of quasi-harmonic sounds~\cite{dolson86pv},
but it is often omitted entirely, or used only in unmodified
reconstruction, as in the \emph{Basic Sinusoidal Model}, described by McAulay
and Quatieri~\cite{mcaulay86speech}.

\begin{figure}
   \centering
   \includegraphics[width=\picwidth]{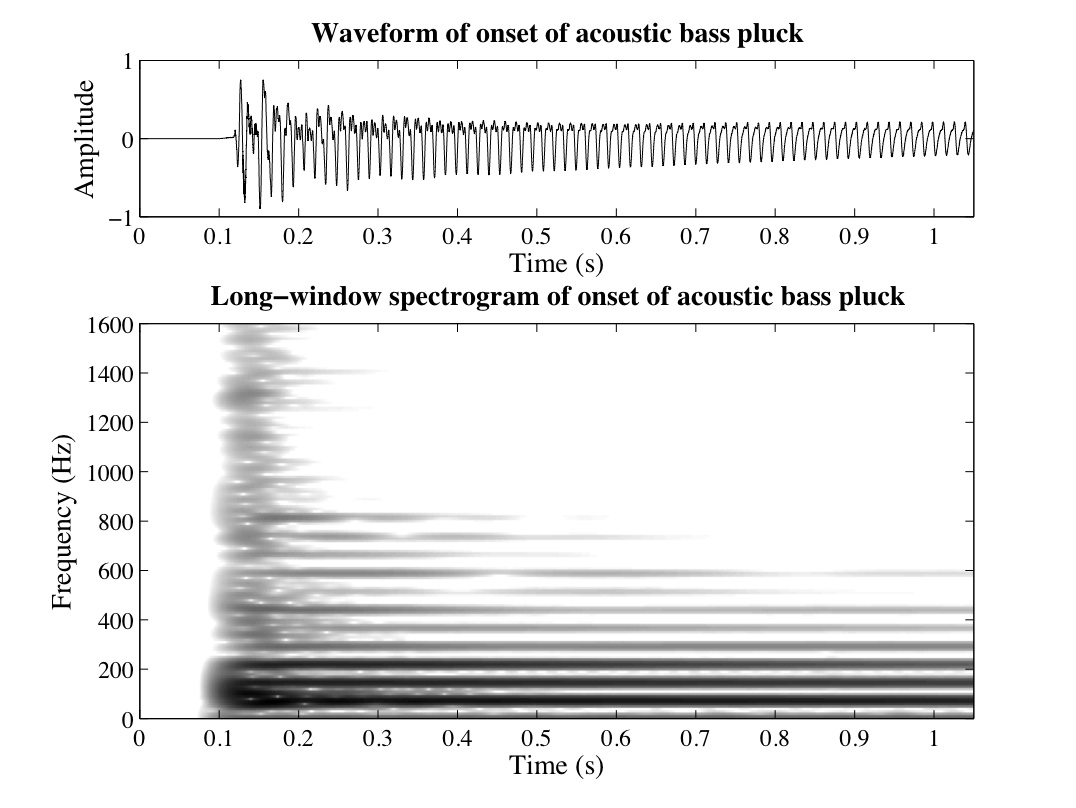} 
   \caption{Spectrogram of an acoustic bass tone having a sharp pluck and a
   fundamental frequency of approximately 73.4~Hz. The spectrogram was computed
   using a  112.7~ms Kaiser window with a shaping
   parameter of~12. The harmonic components are resolved but
   the sharp attack is smeared by the duration of the analysis
   window.
   }
   \label{fig:longpluck}
\end{figure}

\begin{figure}
   \centering
   \includegraphics[width=\picwidth]{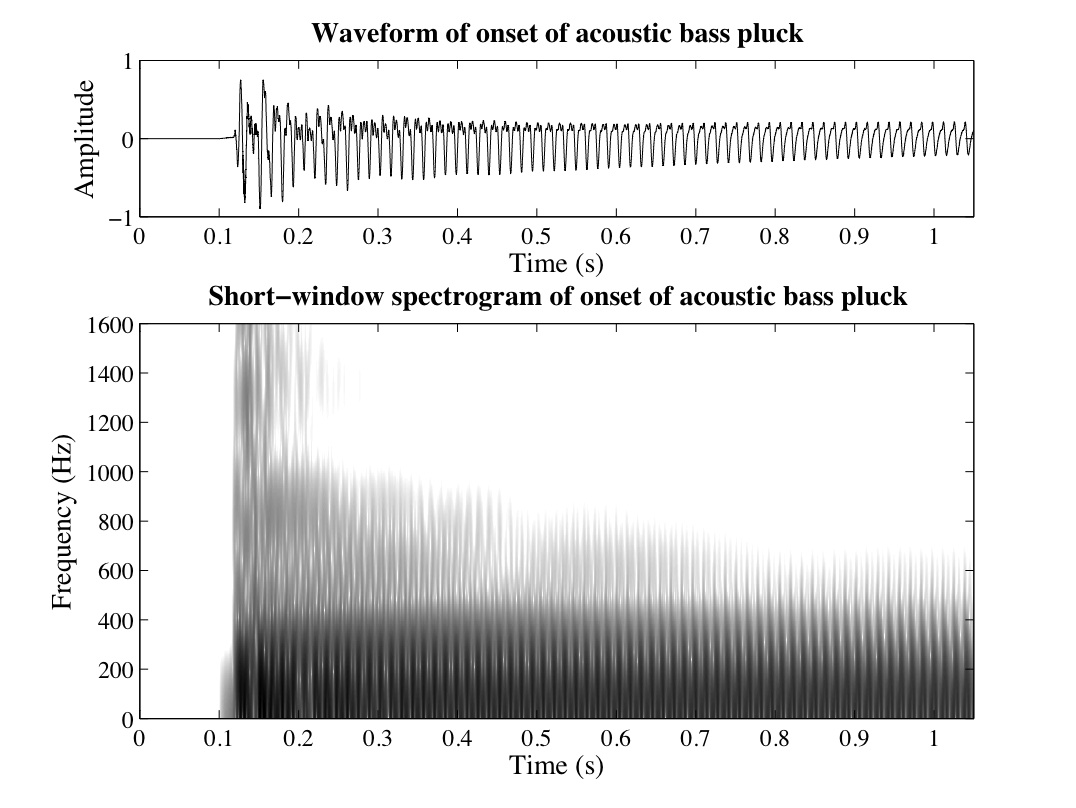} 
   \caption{Spectrogram of an acoustic bass tone having a sharp pluck and a
   fundamental frequency of approximately 73.4~Hz. The spectrogram was computed
   using a  10~ms Kaiser window with a shaping
   parameter of~12. The attack is less distorted than in Figure~\ref{fig:longpluck}
   (though some smearing is still evident)
   but the harmonic structure is no longer discernible.
   }
   \label{fig:shortpluck}
\end{figure}

As a \tfr, the spectrogram has relatively poor resolution.
Time and frequency resolution are governed by the 
choice of analysis window, $h(t)$, and greater concentration
in one domain is accompanied by greater smearing 
in the other. This smearing can be seen in 
Figures~\ref{fig:longpluck} and~\ref{fig:shortpluck}, 
which show spectrograms of a single pluck of an acoustic bass. 
The decaying
part of the tone (after the pluck) is well-represented by nearly-harmonic 
sinusoidal components having a fundamental frequency of approximately
73.4~Hz. 
A very short-duration analysis window is needed
in order to represent the temporal structure of the abrupt attack, but
any window function short enough to provide the necessary temporal precision
is much too wide in frequency to resolve the harmonic components of the tone
(spaced at about 73.4~Hz). In Figure~\ref{fig:longpluck}, the spectrogram
has been computed using a 112.7~ms Kaiser window with a shaping 
parameter of~12. The harmonic components are resolved but
the sharp attack is smeared by the duration of the analysis
window. In Figure~\ref{fig:shortpluck}, the spectrogram
has been computed using a much shorter Kaiser window
(10~ms). The attack is less distorted but the harmonic structure 
is no longer discernible. The combination of poor resolution and 
poor precision often makes it necessary to use two or more
spectrograms, like the two in 
Figures~\ref{fig:longpluck} and~\ref{fig:shortpluck},
to analyze an audio signal.


A \tfr\ having improved resolution, relative to the spectrogram, 
is the \WV\ distribution
\begin{equation}
\label{eqn:wv}
W_{x}(t,\omega) = \int x(t+\tau/2) x^\ast(t-\tau/2)e^{-j\omega \tau}d\tau
\end{equation}
which may be interpreted as a \stft\  with a window function that is perfectly
matched to the signal.
The \WV\ distribution is highly-concentrated in time and frequency, but it is also highly
nonlinear and non-local. Consequently, this distribution is very sensitive to noise, 
and generates cross-components that often mask the components of interest, making
it difficult to extract useful information concerning the distribution of energy in
multi-component signals.

\begin{sloppypar}
Cohen's class of bilinear \tfrs\ ~\cite{cohen95time}
is a class of ``smoothed'' \WV\ distributions, defined
\end{sloppypar}
\begin{equation}
\label{eqn:tfr}
C_{x}(t,\omega) = \iint W_{x}(\tau, \nu) \Phi(\tau -t, \nu -\omega) d\tau d\nu
\end{equation}
where $\Phi(t,\omega)$ is a 
smoothing kernel that can reduce sensitivity to noise and suppresses
cross-components, at the expense of smearing the distribution in time and frequency.
This smearing causes the distribution to be non-zero in regions where the
true \WV\ distribution shows no energy.
$C_{x}(t,\omega)$ can be considered an average 
over a domain centered at the point $t,\omega$
of the values of the \WV\ distribtuion at neighboring points $t-\tau,\omega -\Omega$
weighted by the value of the smoothing kernel $\Phi(\tau,\Omega)$.

The spectrogram in Equation~\ref{eqn:specgram} is a member of Cohen's
class. It is a smoothed \WV\ distribution
with the smoothing kernel equal to the \WV\ distribution of the window function
$h(t)$, that is,
\begin{equation}
\label{equ:specgramkernel}
\Phi(t,\omega) = W_{h}(t,\omega).
\end{equation}

The method of reassignment smoothes the \WV\ distribution, but then refocuses 
the distribution back to the true regions of support of the signal components.
The method has been shown to 
reduce time and frequency smearing of 
any member of Cohen's class~\cite{auger95improving},
but we will focus here on its application to the spectrogram, and by 
extension, to short-time Fourier analysis of time-varying audio signals.
In the case of the reassigned spectrogram, 
the short-time phase spectrum, $\phi(t,\omega)$, is used to correct the nominal
time and frequency coordinates of the spectral data, and map it back
nearer to the true regions of support of the analyzed signal.

\section{The Method of Reassignment}
\label{sec:method}

\begin{sloppypar}
Pioneering work on the method of reassignment was first published by 
Kodera, Gendrin, and de~Villedary under the name of
{\em Modified Moving Window Method}~\cite{kodera78analysis}.
Their technique enhances the resolution in time and frequency 
of the classical {\em Moving Window Method}, the \tfr\ constructed
from the squared magnitude of the \mwt\ defined in Equation~\ref{eqn:mwt},
by assigning to each data point a new
time-frequency coordinate that better-reflects the distribution of
energy in the analyzed signal.
\end{sloppypar}

In the classical moving window method, a  time-domain signal, $x(t)$ is decomposed
into a set of coefficients,
$\epsilon( t, \omega )$, based on a a set of elementary signals, $h_{\omega}(t)$, defined
\begin{equation}
h_{\omega}(t) = h(t) e^{j \omega t} 
\end{equation}
where $h(t)$ is a (real-valued) lowpass kernel function, like
the window function in Equation~\ref{eqn:stft}. The coefficients in this
decomposition are defined
\begin{align}
\epsilon( t, \omega ) 
	&= \int x(\tau) h( t - \tau ) e^{ -j \omega \left[ \tau - t \right]} d\tau \\
	&= e^{ j \omega t}  \int x(\tau) h( t - \tau ) e^{ -j \omega \tau } d\tau \\
	&= e^{ j \omega t} X(t, \omega) \\
	&= X_{t}(\omega)
\end{align}
In other words, the coefficients in the moving window method are computed
from the \mwt\ defined in Equation~\ref{eqn:mwt}. In the moving window
method, the \tfr\ is constructed from the squared magnitude of the coefficients,
and since the magnitude of these coefficients is identical to the magnitude of the
\stft\ coefficients (see Equation~\ref{eqn:Mequal}), this \tfr\ is exactly
equivalent to the spectrogram.

$x(t)$ can be reconstructed from the moving window coefficients by
\begin{align}
x(t)  	& = \iint X_{\tau}(\omega) h^{\ast}_{\omega}(\tau - t) d\omega d\tau \\
  	& = \iint X_{\tau}(\omega) h( \tau - t ) e^{ -j \omega \left[ \tau - t \right]}  d\omega d\tau \\
	&= \iint M_{\tau}(\omega) e^{j \phi_{\tau}(\omega)} h( \tau - t ) e^{ -j \omega \left[ \tau - t \right]}  d\omega d\tau \\
	&= \iint M_{\tau}(\omega) h( \tau - t ) e^{ j \left[ \phi_{\tau}(\omega) - \omega \tau+ \omega t \right] } d\omega d\tau
\end{align}

For signals having magnitude spectra, $M(t,\omega)$, whose
time variation is slow relative to the phase variation, the maximum contribution
to the reconstruction integral comes from the vicinity of the point $t,\omega$
satisfying the phase stationarity condition
\begin{align}
\frac{\partial}{\partial \omega} \left[ \phi_{\tau}(\omega) - \omega \tau +  \omega t\right]  & = 0 \\
\frac{\partial}{\partial \tau} \left[ \phi_{\tau}(\omega) - \omega \tau +  \omega t \right] & = 0 
\end{align}
or equivalently, around the point $\hat{t}, \hat{\omega}$  defined by
\begin{align}
\label{eqn:ratimeop}
\hat{t}(\tau, \omega) & = \tau -  \frac{\partial \phi_{\tau}(\omega)}{\partial \omega} = 
		-  \frac{\partial \phi(\tau, \omega)}{\partial \omega} \\
\label{eqn:rafreqop}
\hat{\omega}(\tau, \omega) & = \frac{\partial \phi_{\tau}(\omega)}{\partial \tau} =
		\omega + \frac{\partial \phi(\tau, \omega)}{\partial \tau} .
\end{align}

This phenomenon has long been known in such fields as optics as the
{\em principle of stationary phase} (see for example~\cite{papoulis68systems}). 
The principle of stationary phase states that for periodic or quasi-periodic
signals, signals that are concentrated in frequency, the variation of the Fourier
phase spectrum not attributable to periodic oscillation is slow with
respect to time in the vicinity of the frequency of oscillation, and in surrounding
regions the variation is relatively rapid. Analogously, for impulsive 
signals, that are concentrated in time, the variation of the phase spectrum
is slow with respect to frequency near the time of the impulse, and in surrounding
regions the variation is relatively rapid. 

In sinusoidal reconstruction, positive and negative contributions to the 
synthesized waveform cancel, due to destructive interference, in frequency regions of
rapid phase variation. Only regions of slow phase variation (stationary
phase) will contribute significantly to the reconstruction, and the
maximum contribution (center of gravity) occurs at the point where the
phase is changing most slowly with respect to time and frequency.

The time-frequency coordinates computed by Equation~\ref{eqn:ratimeop} 
and Equation~\ref{eqn:rafreqop} 
are the local group delay, $\hat{t}_{g}(t,\omega)$, and local instantaneous 
frequency, $\hat{\omega} _{i}(t,\omega)$, and are computed from the phase of the \stft, which is
normally ignored when constructing the spectrogram, though it is known to contain
significant information about the signal. These quantities are ``local''
in the sense that they are represent a windowed and filtered signal
that is localized in time and frequency, and are not global 
properties of the signal under analysis.

The \mmwm\ changes (reassigns) the point of attribution 
of $\epsilon(t,\omega)$ to this point of maximum contribution 
$\hat{t}(t,\omega), \hat{\omega}(t,\omega)$, 
rather than to the point $t,\omega$ at which it is computed. 
This point is sometimes called the ``center of gravity'' of the distribution, by way of 
analogy to a mass distribution (in fact, Kodera~\textit{et al.} demonstrated that 
the coordinates $\hat{t}(t,\omega), \hat{\omega}(t,\omega)$ 
represent the center of gravity of Rihaczek's complex
energy distribution~\cite{rihaczek68signal} for a real signal filtered by the \stft).
This analogy is a useful reminder that
the attribution of spectral energy to the center of gravity of its distribution only makes 
sense when there is energy to attribute, so the method of reassignment has no 
meaning at points where the spectrogram is zero-valued. 

\section{Local Estimation of Instantaneous Frequency}
\label{sec:if}

The group delay, defined in Equation~\ref{eqn:ratimeop},
is often interpreted as the time delay, or average time,
associated with a particular frequency, and its adoption as the reassigned
temporal coordinate is consistent with that interpretation. 
In fact, it can easily be shown that the group delay (or equivalently,
the time reassignment operation) exactly predicts the time of 
an impulse that lies in the region of support of the analysis
window $h(t)$.

The notion of instantaneous frequency also has a long 
history in the signal processing literature, but it is normally 
computed from the {\em signal} phase, rather than the {\em spectral}
phase. Specifically, when a signal is expressed in analytic form, 
\begin{equation}
x(t) = A(t) e^{j \theta (t)}
\end{equation}
where $A(t)$ is the (real) amplitude envelope and $\theta (t)$ is
the (real) phase function, 
then the instantaneous frequency is defined as the
derivative with respect to frequency of the {\em signal} phase, 
$\theta (t)$, that is
\begin{equation}
\omega_{i}(t) = \frac{d \theta (t)}{d t}
\end{equation}


While it is clearly possible to obtain a single-component analytic 
representation of any signal (or, indeed, an infinite number of such
representations), such a representation is not intuitively
satisfying for most audio signals. Pitched sounds, such as 
musical instrument tones,
are characterized by quasi-harmonic spectra, and a representation
as a sum of components representing the various harmonic partials is 
more revealing and intuitive. Vocal sounds are often analyzed as
the response of a resonant system (the vocal tract) to some excitation
signal, and a multicomponent representation that identifies the formants
of the resonant system is more informative.

A crucial insight in the development of the method of frequency 
reassignment follows from the interpretation of the \stft\ as a demodulated bank of
linear time-invariant bandpass filters, wherein each filter,
$h_{\omega}(t)$, has an impulse response determined by
\begin{equation}
\label{eqn:bpf}
h_{\omega}(t) = h(t) e^{j \omega t}
\end{equation}
Since the window function, $h(t)$ is the impulse response of a \fir\ lowpass
filter, the modulated window function is the impulse response of a 
\fir\ bandpass filter with passband centered at frequency $\omega$.
Since $h(t)$ is real, $h_{\omega}(t)$ is complex and, therefore, describes
a filter having an assymmetric frequency response. An analogous bandpass filter
having a real impulse response would pass frequencies around $\pm \omega$, 
but $h_{\omega}(t)$ passes only frequencies around~$\omega$.

In the filterbank interpretation of the \stft, 
$X( t,\omega )$ describes the output of a bank of such filters excited
by the input signal, $x(t)$. That is,
\begin{align}
X(t,\omega) &= \int x(\tau) h(t-\tau)e^{-j\omega \tau}d\tau \\
		&= \int x(\tau) h(t-\tau)e^{j\omega \left[ t - \tau \right]}e^{-j\omega t}d\tau \\
		&= e^{-j\omega t} \int x(\tau) h_{\omega}(t-\tau)d\tau \\		
		&= e^{-j\omega t} \left[ x(t) \ast h_{\omega}(t) \right]
\end{align}

In this interpretation,  $X( t, \omega_{0} )$ describes the 
demodulated output of a single bandpass filter,
centered at frequency $\omega_{0}$. The output of the filter is considered to be a single
complex exponential having magnitude $M(t, \omega_{0})$ and a phase composed of
a linear component, due to sinusoidal oscillation at frequency $\omega_{0}$, and another
time-varying component, $\phi(t, \omega_{0})$, accounting for the deviation from pure
sinusoidal oscillation at frequency $\omega_{0}$. $X( t, \omega_{0} )$ is the output of the 
filter demodulated to remove the sinusoidal oscillation at frequency $\omega_{0}$, and
the output of the \mwt\ $X_{t}(\omega_{0}) = X(t,\omega_{0}) e^{j\omega_{0} t}$,
is the raw output of the bandpass filter centered at frequency $\omega_{0}$.

Kodera~\textit{et al.}~\cite{kodera78analysis} 
showed that the local instantaneous frequency,
computed from the derivative with respect to time of the 
{\em spectral} phase, $\phi(t, \omega)$, is equal to the instantaneous
frequency of the bandpass filtered signal that is the output of the 
\stft\ at the coordinates~$t,\omega$.
For a signal 
\begin{equation}
\label{eqn:single}
x(t) = A(t) e^{j \theta(t)}
\end{equation}
having instantaneous frequency 
\begin{equation}
\omega_{i}(t) = \frac{d \theta(t)}{dt},
\end{equation}
$X(t,\omega_{0}) e^{j\omega_{0} t}$ 
is the output of the bandpass filter
$h_{\omega_{0}}(t)$, centered at frequency $\omega_{0}$,
 when the input is $x(t)$. If, at time $t$, the instantaneous
frequency of the input, $\omega_{i}(t)$ is far from $\omega_{0}$,
such that the instantaneous frequency of the input is outside the passband of the
filter centered at $\omega_{0}$, then the output of the filter is essentially zero.
If, on the other hand, $\omega(t)$, is within the passband of the filter
(near $\omega_{0}$), then the signal passes through the filter unaltered
except for a scale factor, equal to the passband gain, and a
constant time delay, so its instantaneous frequency can be computed
from the phase of the response of $h_{\omega_{0}}(t)$ or any 
filter that passes $x(t)$. 

The filters 
that comprise the \stft\ introduce only a 
constant offset to the phase of any component that they pass,
and the spectral phase obtained from the response of a single
\stft\ filter differs only by a constant offset from the phase of 
a single component whose frequency lies in the passband of
that filter. Therefore, the derivative with respect to time
of the filtered signal, $X_{t}(\omega)$, 
is equal to the derivative with respect to time of the
original signal, $x(t)$, and so the instantaneous frequency of that component
can be computed from the phase of the \stft\ evaluated at $\omega$. That is, 
\begin{align}
\omega_{i}(t)
	&= \frac{\partial}{\partial t}  \arg\{ x(t) \} \\
	&= \frac{\partial }{\partial t} \arg\{ X_{t}(\omega) \} \\
	&= \frac{\partial }{\partial t} \arg\{ e^{j\omega t} X(t,\omega) \} \\
	&= \omega + \frac{\partial \phi(t, \omega)}{\partial t} \\
	&= \hat{\omega}(t, \omega)
\end{align}
for any $\omega$ such that $h_{\omega}(t)$ is
the impulse response of a filter that passes $x(t)$.

The \stft\ filters may also introduce a time delay (if, for example, the frequency
of the signal under analysis is not exactly equal to the center frequency of
the filter), but this delay is precisely the group delay. Therefore, the local 
instantaneous frequency computed from the derivative with respect to time 
of the spectral phase is equal to the instantaneous frequency of the signal 
at a time offset from the center of the analysis window by the group delay, 
computed from the derivative with respect to frequency of the spectral phase.
The \stft\ coefficients evaluated at time $t$ and frequency $\omega$ are
mapped from the geometric center of the analysis window~($t,\omega$) 
onto the region of support of the analyzed signal by the 
reassignment operations in Equation~\ref{eqn:ratimeop} and Equation~\ref{eqn:rafreqop}.

The term {\em local instantaneous frequency} indicates that 
$\hat{\omega}(t, \omega)$ is the instantaneous frequency of the dominant component
at a particular time and frequency. Nelson calls it the {\em channelized
instantaneous frequency}~\cite{nelson02instantaneous,nelson01cross}, 
to emphasize that it is the instantaneous
frequency of a component passing through a single \stft\ channel.
Flanagan also showed that instantaneous frequency 
could be computed for each \stft\ bin from partial derivatives of the phase
spectrum~\cite{flanagan66phase}, and his method has been 
widely used for obtaining estimates of fundamental frequency 
(see for example~\cite{nakatani04robust} and~\cite{goto00robust}).

\begin{figure}
   \centering
   \includegraphics[width=\picwidth]{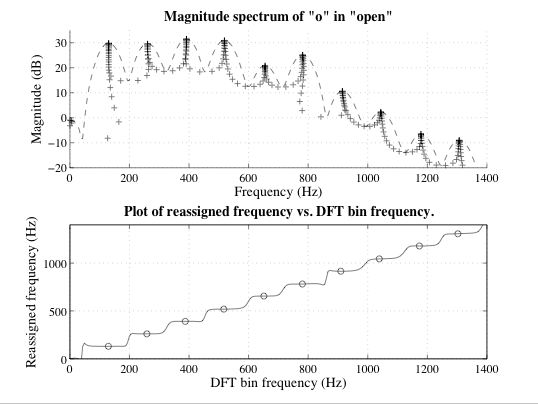} 
   \caption{Demonstration of frequency reassignment in a single spectrum
   for a fragment of speech (the ``o'' in ``open''), computed using a 33.6~ms
   Kaiser analysis window with a shaping parameter of~12. 
   The upper plot
   shows the conventional (dashed lines) and reassigned 
   (crosses) magnitude spectra.
   The lower plot shows the mapping of nominal (\ft\ bin) frequency
   to reassigned frequency for the same fragment of speech. 
   The flat portions of the lower curve represent
   regions in which the energy in many transform bins is reassigned to the same
   frequency. The circled points show the samples having locally minimal 
   frequency reassignments.}
   \label{fig:freqsplot}
\end{figure}

Figure~\ref{fig:freqsplot} demonstrates the effect of frequency
reassignment for a fragment of voiced speech. The upper plot
shows the conventional (dashed lines) and reassigned 
(crosses) magnitude spectra.
The lower plot shows the mapping of nominal (\ft\ bin) frequency
to reassigned frequency for the same fragment of speech.
Near the frequencies of strong harmonics, 
the mapping is flat, as all nearby transform data is reassigned
to the frequency of the dominant harmonic component. 
This \emph{consensus}, or clustering
among reassigned frequency estimates 
in the vicinity of spectral peaks can be used as an indicator
of the reliability of the time-frequency data~\cite{gardner05instantaneous}. 
If the reassigned frequencies for neighboring \stft\ channels are
all very similar, then there is said to be 
a high degree of consensus and the quality of the
frequency estimates is assumed to be good.

The method of time-frequency reassignment has been used in a variety of
applications for obtaining improved time and frequency estimates
for time-varying spectral 
data~\cite{plante98improvement,hainsworth01analysis,fitz01on}.
Often, only the frequency reassignment operation is used to compute 
instantaneous frequency estimates~\cite{nakatani04robust,gardner05instantaneous}.
It should be noted that, for many choices of analysis window,
the much simpler method of parabolic interpolation of the magnitude spectrum, 
proposed by Smith and Serra~\cite{serra90spectral} gives very similar 
\emph{frequency} estimates (and in some cases, even more precise
estimates, according to~\cite{keiler02survey}). 
We are, however, aware of no competing method for 
improving the accuracy of the \emph{time} estimates in short-time 
spectral analysis.

\section{Separability}
\label{sec:separability}
The \stft\ can often be used to estimate the 
amplitudes and phases of the individual components 
in a \emph{multi-component} signal, such as a quasi-harmonic musical instrument
tone (see, for example, \cite{dolson86pv}).
Moreover, the time and frequency reassignment operations described 
by Equation~\ref{eqn:ratimeop} and Equation~\ref{eqn:rafreqop} can be used to 
sharpen the representation by attributing the spectral energy reported by 
the \stft\ to the point that is the local center of gravity of the complex energy 
distribution~\cite{fitz01on}.

For a signal consisting of a single component, 
described by Equation~\ref{eqn:single}, 
the instantaneous frequency 
can be estimated from the partial derivatives of phase of any \stft\
channel that passes the component, as shown above. 
If the signal is to be decomposed
into many components, 
\begin{equation}
\label{eqn:amfm}
x(t) = \sum_{n} A_{n}(t) e^{j \theta_{n}(t)}
\end{equation}
and the instantaneous frequency of each component 
is defined as the derivative of its phase with respect to time, 
that is,
\begin{equation}
\label{eqn:if}
\omega_{n}(t) = \frac{d \theta_{n}(t)}{d t},
\end{equation}
then the instantaneous frequency of each individual component 
can be computed from the phase of the response of a filter that passes
that component, {\em provided that no more than 
one component lies in the passband of the filter.}

This is the property, in the frequency domain, that Nelson
called ``separabilty''~\cite{nelson02instantaneous,nelson01cross}
and we require this property of all the signals we analyze. If this property is not met, then
we cannot achieve the desired multicomponent decomposition, because
we can not estimate the parameters of individual components from the \stft\ and
we must choose a different analysis window so that the separability criterion
is satisfied. 

If the components of a signal are separable in frequency with respect
to a particular short-time spectral analysis window, 
then the output of each \stft\ filter is a filtered version of,
at most, a {\em single} dominant (having significant energy) 
 component, and so the derivative, with
respect to time, of the phase of the $X(t,\omega_{0})$ is equal to the derivative 
with respect to time, of the phase of the dominant component at $\omega_{0}$.
Therefore, if a component, $x_{n}(t)$,
having instantaneous frequency $\omega_{n}(t)$
is the dominant component in the vicinity of $\omega_{0}$, then
the instantaneous frequency of that component
can be computed from the phase of the \stft 
evaluated at $\omega_{0}$. That is, 
\begin{align}
\omega_{n}(t) 
	&= \frac{\partial}{\partial t}  \arg\{ x_{n}(t) \} \\
	&= \frac{\partial }{\partial t} \arg\{ X(t,\omega_{0}) \}
\end{align}
 Thus, the partial derivative with respect to time of the phase of the \stft\
 can be used to compute the instantaneous frequencies of the individual 
 components in a signal described by Equation~\ref{eqn:amfm}, provided only 
 that the components are separable in frequency by the chosen analysis 
 window.

Just as we require that each bandpass filter in the \stft\ filterbank pass at most
a single complex exponential component, we require that two temporal events be sufficiently
separated in time that they do not lie in the same windowed segment of the input
signal. This is the property of separability in the time domain, and
 is equivalent to requiring that the time between two events be greater
than the length of the impulse response of the \stft\ filters, the span of non-zero
samples in $h(t)$. 

Separability in time and in frequency is required of components we wish to resolve
in a reassigned \tfr. If the components in a decomposition 
are separable in time and frequency in a certain \tfr, then the components can be 
resolved by that \tfr, and using the method of reassignment, can be 
characterized with much greater precision than is possible using classical methods.

\begin{figure}
   \centering
   \includegraphics[width=\picwidth]{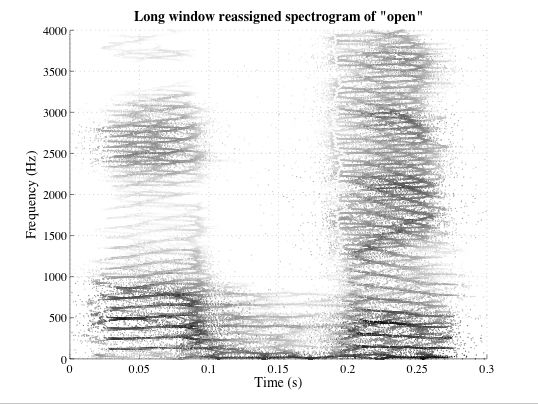} 
   \caption{Long-window reassigned spectrogram of the word ``open'',
   computed using a  54.4~ms Kaiser window with a shaping
   parameter of~9,
   emphasizing harmonics.
   }
   \label{fig:longopen}
\end{figure}

\begin{figure}
   \centering
   \includegraphics[width=\picwidth]{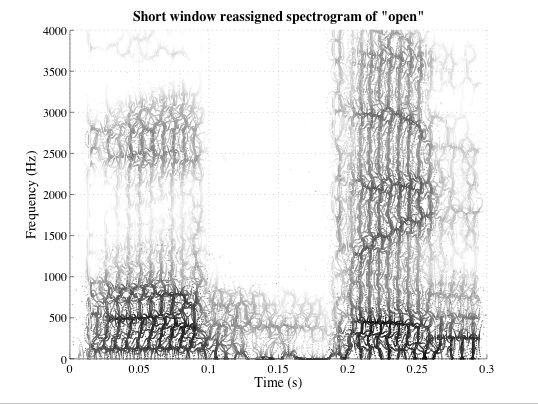} 
   \caption{Short-window reassigned spectrogram of the word ``open'',
   computed using a  13.6~ms Kaiser window with a shaping
   parameter of~9,
   emphasizing formants and glottal pulses.
   }
   \label{fig:shortopen}
\end{figure}

For any signal, there are an infinite number of decompositions 
of the form given in Equation~\ref{eqn:amfm}.
The separability property must be considered in the context of the 
desired decomposition. Figure~\ref{fig:longopen} shows a reassigned
spectrogram of a speech signal computed using an analysis window 
that is long relative to the time between glottal pulses. 
The harmonics are clearly visible, and the 
formant frequencies can also be discerned, but the individual 
glottal pulses are smeared in this representation, because
many pulses are covered by each analysis window
(that is, the individual pulses are not separable, in time,
by the chosen analysis window). 
Figure~\ref{fig:shortopen} shows a
spectrogram of the same speech signal computed
using an analysis window that is much shorter than the
time between glottal pulses. In this representation, the
individual pulses are clearly visible, because no window spans
more than one pulse, but the harmonic frequencies
are not visible, because the main lobe of the analysis window
spectrum is much wider than the spacing between the harmonics
(that is, the harmonics are not separable, in frequency,
by the chosen analysis window). 

\section{Efficient Computation of Reassigned Times and Frequencies}
\label{sec:efficient}

In digital signal processing, it is most common to sample the time and frequency
domains. The \dft\ is used to compute samples $X(k)$ of the \ft\ from samples $x(n)$ of a
time domain signal, 
\begin{equation}
X(k) = \sum_{m=0}^{N-1} x(m) e^{\frac{-j 2 \pi k m}{N}} 
\end{equation}
where $x(n)$ is the time domain signal sampled at $t_{n}=nT$ for sampling period~$T$,
$X(k)$ is the \dft\ coefficients, equal (for adequately-sampled bandlimited signals) 
to samples of the \ft\ at (radian) frequencies $\omega_{k} = 2 \pi k / N$.
For highly-composite $N$, such as powers of two, the \dft\ can be computed 
very efficiently using a fast Fourier transform algorithm.

The discrete \stft\ can be computed 
\begin{align}
X(n,k) 
&= \sum_{m=n-N+1}^{n} x(m) h(n - m) e^{\frac{-j 2 \pi k m}{N}} \\
&= e^{\frac{j 2 \pi k n}{N}}  \sum_{m=0}^{N-1} x(m+n) h(-m) e^{\frac{-j 2 \pi k m}{N}} \\
&= e^{\frac{j 2 \pi k n}{N}}  X_{n}(k)
\end{align}
where $h(n)$ is samples of a real-valued, finite-length 
window function that is non-zero only on the range $n=0 \ldots N-1$, and is
analogous to the analysis window $h(t)$ in Equation~\ref{eqn:stft}.
$X_{n}(k)$ is the \dft\ of a shifted and windowed input signal, 
the discrete time \mwt\ defined in Equation~\ref{eqn:mwt}. 
For a signal sampled with sampling period~$T$ seconds,
the $N$-point discrete \stft\ computes samples of
the \stft\ at times $t_{n}=nT$ and (radian) frequencies 
$\omega_{k}=2 \pi k/N$. 


The reassignment operations proposed by Kodera~\textit{et al.} cannot be
applied to the discrete \stft\ data, because partial derivatives cannot
be computed directly on data that is discrete in time and frequency. 
It has been suggested that this difficulty has been the primary barrier 
to wider use of the method of reassignment~\cite{auger94why}. 

\subsection{Approximation of the Partial Derivatives of Phase}
\label{sec:approx}

It is possible to approximate the partial derivatives using finite differences. 
For example, the phase spectrum can be evaluated at two nearby times,
and the partial derivative with respect to time be approximated
as the difference between the two values divided by the time difference, as in:
\begin{align}
\frac{\partial \phi(t, \omega)}{\partial t} & \approx
	 \frac{1}{\Delta t}  \left[ \phi(t + \frac{\Delta t}{2}, \omega) - \phi(t - \frac{\Delta t}{2}, \omega) \right] \\
\frac{\partial \phi(t, \omega)}{\partial \omega} & \approx
	 \frac{1}{\Delta \omega} 
	  \left[ \phi(t, \omega+ \frac{\Delta \omega}{2}) - \phi(t, \omega-\frac{\Delta \omega}{2}) \right] 
\end{align}

For sufficiently small values of $\Delta t$ and $\Delta \omega$,
this finite-difference method yields 
good approximations to the partial derivatives of phase, 
because in regions of the spectrum in which the evolution of the phase
is dominated by rotation due to sinusoidal oscillation of a single,
nearby component, the phase is a linear function. 

But the phase of the Fourier transform is 
the argument of a complex quantity, and can only be computed modulo $2 \pi$. This 
{\em phase wrapping} effect, that equates all phases differing by a multiple of $2 \pi$,
has no practical impact on the individual phase values, but
is of some consequence when two phases are combined as in the finite difference
derivative approximation, because the difference between two phase values may not
be preserved by the wrapping process. 

Fortunately, $\Delta t$ and $\Delta \omega$, can be chosen to be
small enough that, for a properly-sampled signal,
the phase difference can be easily ``unwrapped''.
The absolute value of the difference in phase between consecutive samples, in
time or in frequency, of the \stft\ cannot exceed~$\pi$ in any region of the
spectrum dominated by a significant oscillating component. Any phase difference
that exceeds~$\pi$, in the absolute sense, has been corrupted by phase wrapping,
and can be corrected by simply adding or subtracting~$2 \pi$ to obtain an
absolute value that is less than~$\pi$. Thus, if $\Delta t$ and 
$\Delta \omega$ are chosen to be one sample, then the finite difference method
can be used to accurately-estimate the reassigned
times and frequencies in regions of the 
spectrum in which the evolution of the phase
is dominated by rotation due to sinusoidal oscillation of a nearby component. 

Charpentier~\cite{charpentier86pitch} 
used a finite difference approximation to Flanagan's
instantaneous frequency equation~\cite{flanagan66phase} and showed
that the approximation could be computed using only a single \ft\
for the case of the Hann analysis window.

Unaware of the work of Kodera~\textit{et al.}, Nelson arrived at a similar method for
computing reassigned time-frequency coordinates for short-time
spectral data  from partial derivatives of the short-time phase
spectrum~\cite{nelson02instantaneous,nelson01cross}. 
Instead of directly computing the first differences of phase, 
Nelson first computes two so-called {\em cross spectral surfaces}, 
\begin{align}
C(t,\omega) &= X(t + \frac{\Delta t}{2}, \omega) X^{\ast}(t - \frac{\Delta t}{2}, \omega)\\
L(t,\omega) &= X(t, \omega+ \frac{\Delta \omega}{2}) X^{\ast}(t, \omega-\frac{\Delta \omega}{2})
\end{align}
The partial derivatives are then approximated by the phase of these cross spectra.
\begin{align}
\frac{\partial \phi(t, \omega)}{\partial t} & \approx
	\frac{1}{\Delta t} \arg( C(t,\omega) ) \\
\frac{\partial \phi(t, \omega)}{\partial \omega} & \approx
	 \frac{1}{\Delta \omega} \arg( L(t,\omega) )
\end{align}

It is easily shown that approximation of the derivatives by means of a
cross spectral surface is equivalent to computing the finite differences directly, only
the differences are unwrapped automatically when the argument
is computed, for example:
\begin{align}
\frac{1}{\Delta t} \arg( C(t,\omega) )
&= 	\frac{1}{\Delta t} \arg \left( X(t + \frac{\Delta t}{2}, \omega) 
					X^{\ast}(t - \frac{\Delta t}{2}, \omega) \right) \\
&= 	\frac{1}{\Delta t} \arg \left( 	
	M(t + \frac{\Delta t}{2}, \omega) e^{j \phi (t + \frac{\Delta t}{2}, \omega) }
	M(t - \frac{\Delta t}{2}, \omega) e^{-j \phi (t - \frac{\Delta t}{2}, \omega) } \right) \\
&= \frac{1}{\Delta t} \arg \left( 	
	M(t + \frac{\Delta t}{2}, \omega) M(t - \frac{\Delta t}{2}, \omega) 
	e^{j \left[ \phi (t + \frac{\Delta t}{2}, \omega) - \phi (t - \frac{\Delta t}{2}, \omega) \right] }
	\right) \\
&= \frac{1}{\Delta t}  \left[ \phi (t + \frac{\Delta t}{2}, \omega) - \phi (t - \frac{\Delta t}{2}, \omega) \right] 
\end{align}

While these linear differences only approximate the partial derivatives of the phase
with respect to time and frequency, they give very good approximations
in regions of the spectrum dominated by a single, significant concentration
of energy, such as an impulse or a sinusoid, because in these regions
the evolution of the phase spectrum is linear in time and in frequency.
In regions of the spectrum having no significant concentration of energy,
the finite difference approximation may not be a good one, but it makes little
sense to compute reassigned time and frequency coordinates when there is no
energy to reassign.
	
\subsection{Evaluation of the Partial Derivatives of Phase Using Transforms}
\label{sec:eval}

Auger and Flandrin~\cite{auger95improving} showed how the method of reassignment, proposed
in the context of the spectrogram by Kodera~\textit{et al.}, could be extended to 
any member of Cohen's class of \tfrs\ by generalizing the 
reassignment operations in Equation~\ref{eqn:ratimeop} and Equation~\ref{eqn:rafreqop} to
\begin{align}
\hat{t} (t,\omega) & = t - 
	\frac{\iint \tau \cdot W_{x}(t-\tau,\omega -\nu) \cdot \Phi(\tau,\nu) d\tau d\nu}
		{\iint W_{x}(t-\tau,\omega -\nu) \cdot \Phi(\tau,\nu) d\tau d\nu } \\
\hat{\omega} (t,\omega) & = \omega - 
	\frac{\iint \nu \cdot W_{x}(t-\tau,\omega -\nu) \cdot \Phi(\tau,\nu) d\tau d\nu}
		{\iint W_{x}(t-\tau,\omega -\nu) \cdot \Phi(\tau,\nu) d\tau d\nu}
\end{align}
where $W_{x}(t,\omega)$ is the \WV\ distribution of $x(t)$, and 
$\Phi(t,\omega)$ is the kernel function that defines the distribution.
They further described an efficient 
method for computing the times and frequencies for the
reassigned spectrogram
efficiently and accurately without explicitly computing the
partial derivatives of phase. 

In the case of the spectrogram, $S_{x}(t,\omega) = | X(t,\omega) |^2$,
the reassignment operations in Equation~\ref{eqn:ratimeop} and Equation~\ref{eqn:rafreqop}
can be computed by
\begin{align}
\label{eqn:aftcorrection}
\hat{t} (t,\omega) & = t - \Re \Bigg\{ \frac{ X_{\mathcal{T}h}(t,\omega) \cdot X^\ast(t,\omega) }
								{ | X(t,\omega) |^2 } \Bigg\}  \\
\label{eqn:affcorrection}
\hat{\omega}(t,\omega) & = \omega + \Im \Bigg\{ \frac{ X_{\mathcal{D}h}(t,\omega) \cdot X^\ast(t,\omega) }
								{ | X(t,\omega) |^2 } \Bigg\}  
\end{align}
where $X(t,\omega)$ is the \stft\ computed
using an analysis window $h(t)$, $X_{\mathcal{T}h}(t,\omega)$ is the \stft\
computed using a time-weighted anlaysis window $h_{\mathcal{T}}(t) = t \cdot h(t)$
and $X_{\mathcal{D}h}(t,\omega)$ is the \stft\ computed using a time-derivative
analysis window $h_{\mathcal{D}}(t) = \frac{d}{dt}h(t)$. 

In this method, using the auxiliary window functions $h_{\mathcal{T}}(t)$ and
$h_{\mathcal{D}}(t)$, the reassignment operations can be computed at
any time-frequency coordinate $t,\omega$ from an algebraic combination of
the values of three Fourier transforms evaluated at $t,\omega$, without
directly evaluating or approximating the partial derivatives of phase.
A method of computing instantaneous frequency equivalent 
to Equation~\ref{eqn:affcorrection}
was independently discovered by Abe~\cite{abe96robust},
and is sometimes used in fundamental frequency estimation
(see for example~\cite{nakatani04robust}).
Since the these algorithms operate only on spectral data
evaluated at a single time and frequency, and do not explicitly 
compute any derivatives, they can easily be implemented
in digital systems using discrete times and frequencies.

The time-weighted window function, $h_{\mathcal{T}}(t)$, 
is trivially computed by pointwise multiplication of the original
window function, $h(t)$, by a time ramp. If the derivative of the
window function is unknown, then $h_{\mathcal{D}}(t)$ can also
be computed numerically.
The derivative theorem for Fourier transforms, which states that
if $X(\omega)$ is the \ft\ of $x(t)$, then the \ft\ of $\frac{d}{dt}x(t)$ is
$j \omega X(\omega)$. That is, 
\begin{equation}
x(t)  \leftrightarrow X(\omega)
\end{equation}
implies
\begin{equation}
\frac{d}{dt}x(t)  \leftrightarrow j \omega X(\omega)
\end{equation}
We can therefore construct the time-derivative window used in the evaluation of the
frequency reassignment operator by computing the Fourier transform of
$h(t)$, multiplying by $j\omega$, and inverting the Fourier transform. That is,
\begin{align}
\frac{d}{dt}h(t) & = \mathrm{FFT}^{-1} \{ j \omega H(\omega) \} \\
& = - \Im \Big\{ \mathrm{FFT}^{-1} \{ \omega H(\omega) \} \Big\}  
\end{align}
and so, in discrete time, 
\begin{equation}
h_{\mathcal{D}}(n) 
= - \Im \Big\{ \mathrm{FFT}^{-1} \big \{ \frac{2 \pi k}{N} H(k) \big \} \Big\}  
\end{equation}

\begin{figure}
\centerline {
\includegraphics[width=\picwidth]{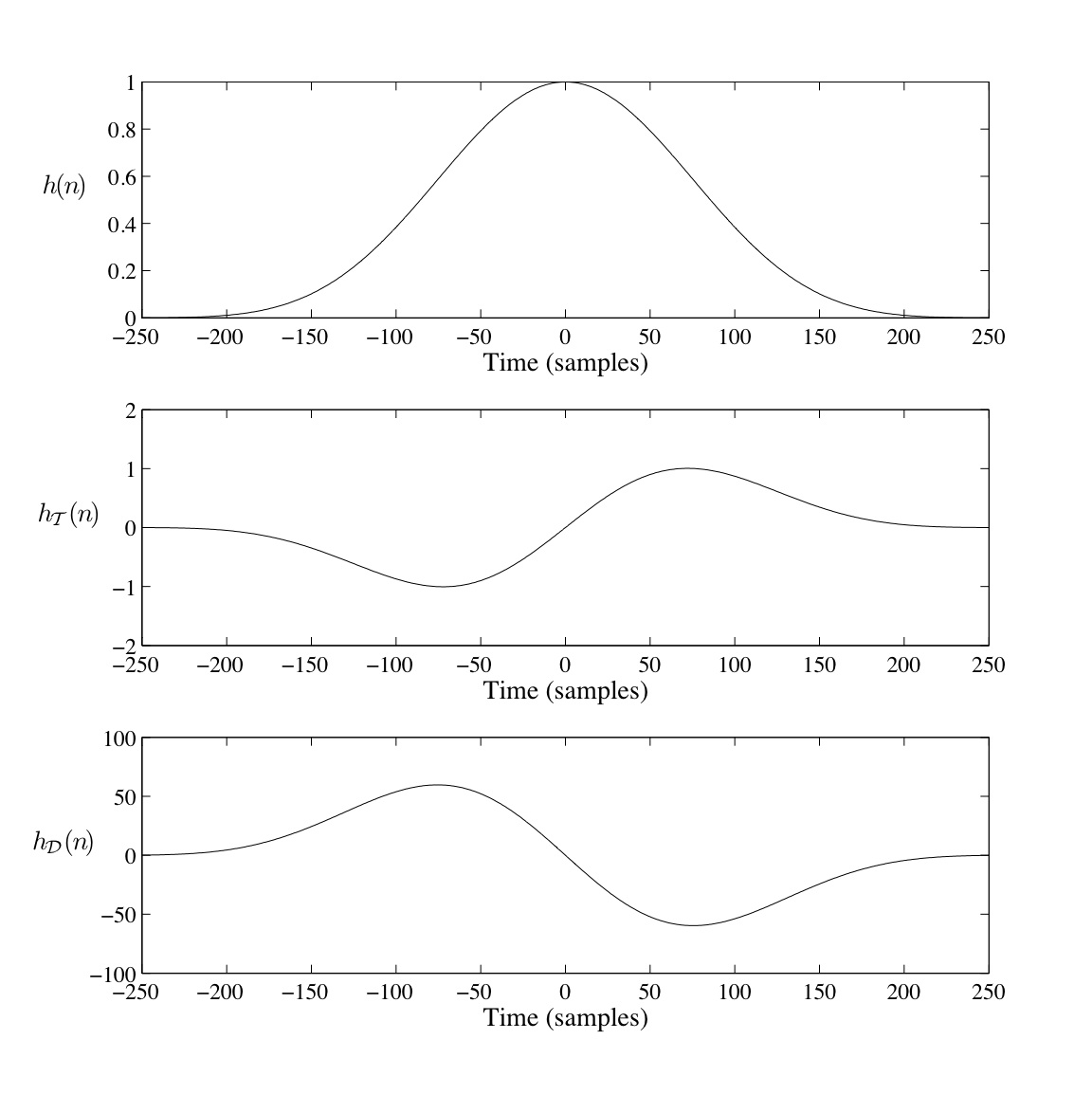}
}
\caption{Representative analysis windows employed in 
the three short-time transforms used to
compute reassigned times and frequencies. Waveform (a) is the original
window function, $h(n)$ (a Kaiser window
with shaping parameter 12.0, in this case), waveform (b) is the
time-weighted window function, $h_{\mathcal{T}}(n)$, 
and waveform (c)
is the frequency-weighted window function,
$h_{\mathcal{D}}(n)$. 
In each case, the plots were made from 501~samples
of the corresponding window function at 44.1~kHz.
\label{fig:RAwindows}}
\end{figure}

The auxiliary short-time analysis windows employed in the computation of
Auger and Flandrin's reassignment operations are shown in 
Figure~\ref{fig:RAwindows} for the case of $h(n)$ being 501~samples
at 44.1~kHz of a Kaiser
window with shaping parameter equal to~12. 

%
%

Using Equation~\ref{eqn:aftcorrection} and Equation~\ref{eqn:affcorrection},
the value of the reassignment operations at $t,\omega$
can be computed from the values of the three \stfts\ at 
$t,\omega$, without direct evaluation of any partial derivatives. 
So if the values of the transforms at discrete time and frequency coordinates, 
$t_{n},\omega_{k}$, are known, then the values of the reassignment operations
can be computed for those discrete times and frequencies without resorting
to discrete approximations to the partial derivatives 
in Equation~\ref{eqn:ratimeop} and Equation~\ref{eqn:rafreqop}. Since the discrete \stft\
computes the values of the \stft\ at discrete times and frequencies,
values of the reassignment operations, $\hat{\omega} (t_{n},\omega_{k})$
and $\hat{t}(t_{n},\omega_{k})$ can be computed from three
discrete \stfts. This gives an efficient method of computing the 
reassigned discrete \stft\ provided only that the $| X(t,\omega) |^2$ is non-zero.
This is not much of a restriction, 
since the reassignment operation itself implies that there is some energy to reassign,
and has no meaning when the distribution is zero-valued.

In the two sections that follow, we derive the reassignment operations
proposed by Auger and Flandrin. 
The implementation of the reassignment operations has been described
elsewhere~\cite{flandrin03time,fulop05algorithms}, and
partial derivations have been 
presented (see, for example~\cite{auger95improving} and~\cite{auger94why}),
but we are not aware of a complete, published derivation.
We include the derivations here for completeness, and because
we have found them to be a useful starting point for 
deriving methods of computing other spectral derivatives.

The procedure in each derivation is the 
same. First, we show that the partial derivative of the spectral phase 
can be expressed in terms of the \stft\ and its
partial derivative. Then, we show that the partial
derivative of the \stft\ can be computed from the transform itself
and a second transform computed using a different window
function. Finally, we combine the results to obtain an 
expression for the reassigned coordinate that does not
include any explicit partial derivatives. 

Readers not interested in the mathematical derivation of the 
reassignment operations can skip ahead to Section~\ref{sec:pruning}.


\subsubsection{Derivation of Efficient Spectrogram Time Reassignment Operator}
\label{sec:derivetime}

In this section, we present the mathematical derivation of the efficient
time reassignment operator discovered by Auger and Flandrin. In 
Section~\ref{sec:derivefreq} we will present the derivation of the 
frequency reassignment operator. 
We begin by restating the time reassignment operation identified by Kodera
\textit{et. al.}, and presented here in Equation~\ref{eqn:ratimeop}
\begin{equation*}
\hat{t}(t,\omega) = -  \frac{\partial \phi(t, \omega)}{\partial \omega}.
\end{equation*}

To arrive at an expression for the partial derivative of spectral phase with respect
to frequency, we first take the partial derivative of $X(t,\omega)$ with respect 
to frequency. 
Applying the product rule of differential calculus to Equation~\ref{eqn:Mphi},
\begin{align}
\frac{\partial}{\partial \omega} X(t,\omega)
	& = \frac{\partial}{\partial \omega} \left[M(t,\omega) e^{j \phi(t,\omega)} \right]  \\
	& =  \frac{\partial M(t,\omega)}{\partial \omega} \cdot e^{j \phi(t,\omega)} 
		+ M(t,\omega) \cdot j  \frac{\partial \phi(t,\omega)}{\partial \omega} 
		e^{j \phi(t,\omega)}  \\
	\label{eqn:freqderivsplit}
	& =  \frac{\partial M(t,\omega)}{\partial \omega} \cdot e^{j \phi(t,\omega)} 
		+ j  \frac{\partial \phi(t,\omega)}{\partial \omega} X(t,\omega) 
\end{align}
\begin{sloppypar}
In order to isolate the partial derivative of phase, we can multiply
by $j X^\ast(t,\omega) / | X(t,\omega) |^2$ (this operation is only
valid when $| X(t,\omega) |^2$ is non-zero, but as noted earlier, the reassignment
operation itself has no meaning when the distribution is zero-valued) 
and simplify to obtain
\end{sloppypar}
\begin{align}
\left[\frac{\partial}{\partial \omega} X(t,\omega)\right] \cdot \frac{j X^\ast(t,\omega)}{| X(t,\omega) |^2}
	& =  \left[  \frac{\partial M(t,\omega)}{\partial \omega} \cdot e^{j \phi(t,\omega)} 
		+ j  \frac{\partial \phi(t,\omega)}{\partial \omega} \cdot X(t,\omega)  \right] 
		\cdot \frac{j X^\ast(t,\omega)}{| X(t,\omega) |^2}  \\	
	& =   \frac{\partial M(t,\omega)}{\partial \omega} \cdot e^{j \phi(t,\omega)}
		 \cdot \frac{j X^\ast(t,\omega)}{| X(t,\omega) |^2}
		-  \frac{\partial \phi(t,\omega)}{\partial \omega} 
		\cdot \frac{| X(t,\omega) |^2}{| X(t,\omega) |^2}  \\
	& =   \frac{\partial M(t,\omega)}{\partial \omega} 
		 \cdot \frac{j M(t,\omega)}{| X(t,\omega) |^2}
		-  \frac{\partial \phi(t,\omega)}{\partial \omega}  \\
	& =  -  \frac{\partial \phi(t,\omega)}{\partial \omega} 
		+ j \cdot \left[\frac{M(t,\omega)}{| X(t,\omega) |^2} \right] \cdot
		 \frac{\partial M(t,\omega)}{\partial \omega} 
\end{align}
Since $M(t,\omega)$ is real-valued,
the real part of this expression is precisely the negative partial derivative 
with respect to frequency of the phase of the \stft.
Thus, we conclude that the partial derivative of spectral phase with 
respect to frequency, and hence, the reassigned time, $\hat{t}$, can
be computed by 
\begin{equation}
\label{eqn:isolatepartfreq}
\hat{t}(t,\omega) = -  \frac{\partial \phi(t, \omega)}{\partial \omega}
	= \Re \Bigg \{ \frac{\partial X(t,\omega)}{\partial \omega} \cdot 
			\frac{j X^\ast(t,\omega)}{| X(t,\omega) |^2} \Bigg \}
\end{equation}

Equation~\ref{eqn:isolatepartfreq} expresses the partial derivative
of the \stft\ phase with respect to frequency in terms of the transform itself 
and its partial derivative with respect to frequency. 

Next, we will show that the partial derivative of the 
\stft\ with respect to frequency can be computed without
explicitly computing or approximating any derivatives.
Taking the partial derivative of the 
\stft\ given by Equation~\ref{eqn:stft}, 
\begin{align}
\frac{\partial}{\partial \omega} X(t,\omega)
	& = \frac{\partial}{\partial \omega} \int x(\tau) h(t-\tau)e^{-j\omega \tau}d\tau  \\
	& = \int x(\tau) h(t-\tau) \left[\frac{\partial}{\partial \omega} e^{-j\omega \tau}  \right] d\tau  \\
	& = \int x(\tau) h(t-\tau) \left[ -j \tau e^{-j\omega \tau}  \right] d\tau  \\
	\label{eqn:freqderivintegral}
	& = -j \int x(\tau) \cdot \tau \cdot h(t-\tau) e^{-j\omega \tau} d\tau  \\
	& = -j t X(t,\omega) + j t X(t,\omega) -j \int x(\tau) \cdot \tau \cdot 
		h(t-\tau) e^{-j\omega \tau} d\tau  \\
	& = -j t X(t,\omega) 
		 + j \int x(\tau) \cdot (t - \tau) \cdot h(t-\tau)e^{-j\omega \tau}d\tau  \\
	& = -j t X(t,\omega) 
		 + j \int x(\tau) h_{\mathcal{T}}(t-\tau)e^{-j\omega \tau}d\tau   \\
	\label{eqn:freqderivnotate}
	& =  -j t X(t,\omega) 
		 + j X_{\mathcal{T}h}(t,\omega) 
\end{align}
where $h_{\mathcal{T}}(t) = t \cdot h(t)$ is the time-weighted anlaysis window
described in Section~\ref{sec:efficient} and shown in Figure~\ref{fig:RAwindows},
and $X_{\mathcal{T}h}(t,\omega)$ is the \stft\ computed using this time-weighted
analysis window.
Multiplying, as before (and with the same caveat concerning zero-valued
distributions), by $j X^\ast(t,\omega) / | X(t,\omega) |^2$, we obtain
\begin{align}
\left[\frac{\partial}{\partial \omega} X(t,\omega)\right] 
		\cdot \frac{j X^\ast(t,\omega)}{| X(t,\omega) |^2} 
	& = \left[-j t X(t,\omega) + j X_{\mathcal{T}h}(t,\omega) \right] \cdot 
		\frac{j X^\ast(t,\omega)}{| X(t,\omega) |^2}  \\
	& =  t \cdot \frac{| X(t,\omega) |^2}{| X(t,\omega) |^2} +
		 \frac{j X_{\mathcal{T}h}(t,\omega) \cdot j X^\ast(t,\omega)}{| X(t,\omega) |^2}  \\
	& =  t - \frac{X_{\mathcal{T}h}(t,\omega) \cdot X^\ast(t,\omega)}{| X(t,\omega) |^2} 
\end{align}

Substituting this expression into Equation~\ref{eqn:isolatepartfreq}, we obtain
\begin{equation}
\hat{t}(t, \omega) = -  \frac{\partial \phi(t, \omega)}{\partial \omega}
\label{eqn:AFratimeop}
	=  t - \Re \Bigg \{ \frac{X_{\mathcal{T}h}(t,\omega) \cdot 
		X^\ast(t,\omega)}{| X(t,\omega) |^2} \Bigg \} 
\end{equation}
which is the time reassignment operation proposed by Auger and Flandrin, 
previously presented in Equation~\ref{eqn:ratimeop}.

\subsubsection{Derivation of Efficient Spectrogram Frequency Reassignment Operator}
\label{sec:derivefreq}

In this section, we present the mathematical derivation of the efficient
frequency reassignment operator discovered by Auger and Flandrin. 
We begin by restating the frequency reassignment operation identified by Kodera
\textit{et. al.}, and presented here in Equation~\ref{eqn:rafreqop}
\begin{equation*}
\hat{\omega}(t, \omega) = \omega + \frac{\partial \phi(t, \omega)}{\partial t} .
\end{equation*}

To arrive at an expression for the partial derivative of spectral phase with respect
to time, we first take the partial derivative of $X(t,\omega)$ with respect 
to time. 
Applying the product rule of differential calculus to Equation~\ref{eqn:Mphi},
\begin{align}
\frac{\partial}{\partial t} X(t,\omega)
	& = \frac{\partial}{\partial t} \left[M(t,\omega) e^{j \phi(t,\omega)} \right]  \\
	& =  \frac{\partial M(t,\omega)}{\partial t}\cdot e^{j \phi(t,\omega)} 
		+ M(t,\omega) \cdot j \frac{\partial \phi(t,\omega)}{\partial t}
		e^{j \phi(t,\omega)}  \\
	& =  \frac{\partial M(t,\omega)}{\partial t} \cdot e^{j \phi(t,\omega)} 
		+ j \frac{\partial \phi(t,\omega)}{\partial t} \cdot X(t,\omega) 
\end{align}
\begin{sloppypar}
In order to isolate the partial derivative of phase, we can multiply
by $X^\ast(t,\omega) / | X(t,\omega) |^2$ (this operation is only
valid when $| X(t,\omega) |^2$ is non-zero, but as noted earlier, the reassignment
operation itself has no meaning when the distribution is zero-valued) 
and simplifying to obtain
\end{sloppypar}
\begin{align}
\frac{\partial X(t,\omega)}{\partial t} \cdot 
	\frac{X^\ast(t,\omega)}{| X(t,\omega) |^2}
	& =  \left[  \frac{\partial M(t,\omega)}{\partial t} \cdot e^{j \phi(t,\omega)} 
		+ j  \frac{\partial \phi(t,\omega)}{\partial t} \cdot X(t,\omega)  \right] 
		\cdot \frac{X^\ast(t,\omega)}{| X(t,\omega) |^2} \\	
	& =  \frac{\partial M(t,\omega)}{\partial t} \cdot e^{j \phi(t,\omega)} \cdot 
		\frac{X^\ast(t,\omega) }{| X(t,\omega) |^2}
		+ j \frac{\partial \phi(t,\omega)}{\partial t} \cdot
		\frac{| X(t,\omega) |^2}{| X(t,\omega) |^2}  \\
	 & =  \frac{\partial M(t,\omega)}{\partial t} \cdot 
		\frac{M(t,\omega) }{| X(t,\omega) |^2}
		+ j \frac{\partial \phi(t,\omega)}{\partial t}
\end{align}
Since $M(t,\omega)$ is real-valued,
the imaginary part of this expression is precisely the partial derivative 
with respect to time of the phase of the \stft.
Thus, we conclude that the value of the frequency reassignment operator can
be computed by 
\begin{equation}
\label{eqn:isolateparttime}
\hat{\omega}(t, \omega) = \omega + \frac{\partial \phi(t, \omega)}{\partial t} 
	= \omega + \Im \Bigg \{ \frac{\partial X(t,\omega)}{\partial t} \cdot 
			\frac{X^\ast(t,\omega)}{| X(t,\omega) |^2} \Bigg \}
\end{equation}

Equation~\ref{eqn:isolateparttime} expresses the partial derivative
of the \stft\ phase with respect to time in terms of the transform itself 
and its partial derivative with respect to time. 

Next, we will show that the partial derivative of the 
\stft\ with respect to time can be computed without
explicitly computing or approximating any derivatives.
Taking the partial derivative of the 
\stft\ given by Equation~\ref{eqn:stft}, 
\begin{align}
\frac{\partial}{\partial t} X(t,\omega)
	& = \frac{\partial}{\partial t} \int x(\tau) h(t-\tau)e^{-j\omega \tau}d\tau  \\
	& = \int x(\tau) \left[\frac{\partial}{\partial t} h(t-\tau) \right] e^{-j\omega \tau}d\tau  \\
	& = \int x(\tau) h_{\mathcal{D}}(t-\tau)e^{-j\omega \tau}d\tau  \\
	& = X_{\mathcal{D}h}(t,\omega)
\end{align}
where $h_{\mathcal{D}}(t) = \frac{d}{dt}h(t)$ is the time-derivative anlaysis window
described in Section~\ref{sec:efficient} and shown in Figure~\ref{fig:RAwindows},
and $X_{\mathcal{D}h}(t,\omega)$ is the \stft\ computed using this time-derivative
analysis window.
Multiplying, as before (and with the same caveat concerning zero-valued
distributions), by $j X^\ast(t,\omega) / | X(t,\omega) |^2$, we obtain
\begin{equation}
\frac{\partial X(t,\omega)}{\partial t} \cdot \frac{X^\ast(t,\omega)}{| X(t,\omega) |^2}
	= 
\frac{X_{\mathcal{D}h}(t,\omega) \cdot X^\ast(t,\omega) }{| X(t,\omega) |^2}
\end{equation}

Substituting this expression into Equation~\ref{eqn:isolateparttime}, we obtain
\begin{equation}
\hat{\omega}(t, \omega) = \omega + \frac{\partial \phi(t, \omega)}{\partial t} 
	= \omega + \Im \Bigg \{ \frac{X_{\mathcal{D}h}(t,\omega) \cdot X^\ast(t,\omega) }{| X(t,\omega) |^2} \Bigg \}
\end{equation}
which is the frequency reassignment operation proposed by Auger and Flandrin, 
previously presented in Equation~\ref{eqn:rafreqop}.




\section{Pruning Reassigned Data to Improve Spectrogram Readability}
\label{sec:pruning}

\begin{figure}
   \centering
   \includegraphics[width=\picwidth]{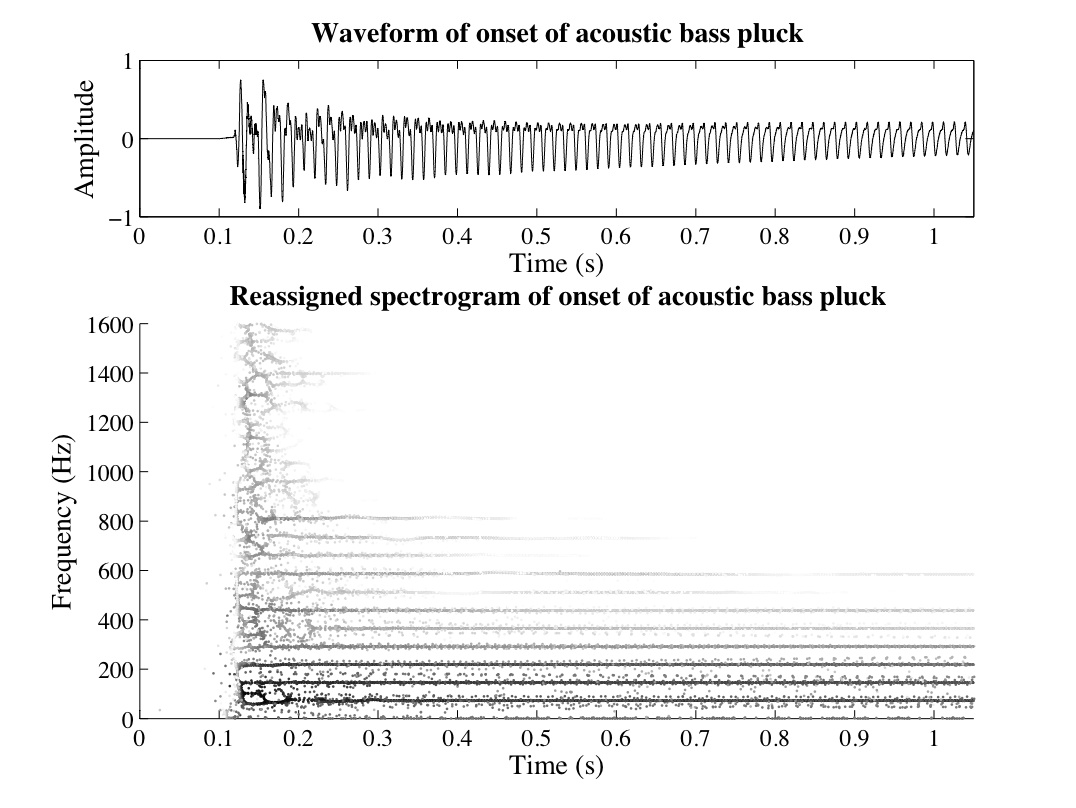} 
   \caption{Reassigned spectrogram for the onset 
   of an acoustic bass tone having a sharp pluck and a
   fundamental frequency of approximately 73.4~Hz. 
   The spectrogram was computed
   using a  65.7~ms Kaiser window with a shaping
   parameter of~12. 
   The sharp attack and the harmonic components are 
   clearly visible.
   }
   \label{fig:rapluck}
\end{figure}

Reassigned time and frequency
estimates are much more precise than those obtained from 
traditional methods, so for many applications, the ``readability''
or ``interpretability'' of the spectral data is much improved
by reassignment. 
The resolving power of the reassigned \stft\
is no greater than that of the classical \stft. The separability conditions
discussed in Section~\ref{sec:separability} apply equally to reassigned and
non-reassigned spectra, and components smeared together
by the analysis window will be smeared in both reassigned and non-reassigned
spectral data. Provided the separability conditions are satisfied, however, 
reassigned spectrograms offer improved clarity in the representation
of quasi-sinusoidal components and improved localization of impulsive
events
This improved readability is evident in the
plot in Figure~\ref{fig:rapluck}, showing a
reassigned spectrogram for the same bass pluck 
that was plotted in Figures~\ref{fig:longpluck} and~\ref{fig:shortpluck}.
The sharp attack is clearly visible in this reassigned data, as are the harmonic
components. 

\begin{figure}
   \centering
   \includegraphics[width=\picwidth]{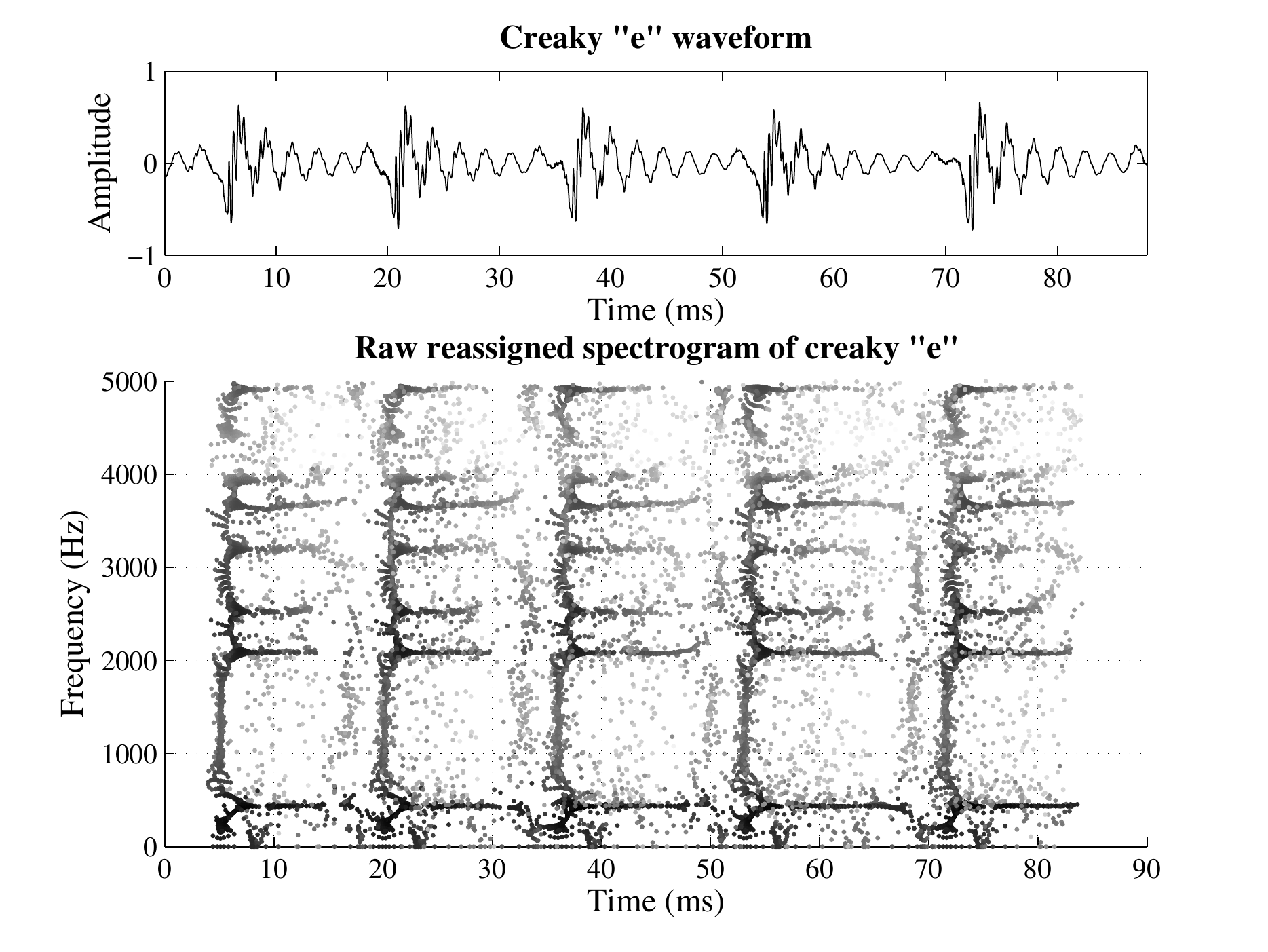} 
   \caption{Reassigned spectrogram of the vowel \emph{e} (day)
   computed using a 7.8~ms Hann window.}
   \label{fig:creakyraw}
\end{figure}

In spite of the obvious gains in clarity, reassigned spectrograms can be 
disappointingly noisy. Seemingly random speckle is visible in regions where 
the reassigned data is not clearly associated with either a sinusoidal component 
or an impulsive component. This random speckle can be seen 
between the harmonics in Figure~\ref{fig:rapluck}, and in the
reassigned spectrogram shown in Figure~\ref{fig:creakyraw}, 
representing a portion of the vowel \emph{e} (d\textbf{a}y), spoken 
in a ``creaky'' (low airflow) voice.
The main glottal impulses appear as dark grey lines, and these result from 
the sudden release of a puff of air into the mouth from below the vocal cords.
There are also fainter secondary impulses of unknown cause (these have 
been attributed to the mechanics of the vocal cord action), appearing just 
prior to the main impulses in this particular voice sample.

Reassignment provides a mapping
from the geometrical center of the short-time analysis window to the center
of gravity of a nearby dominant spectral component, but in low-energy regions of 
the spectrum, where there is no dominant component, data is reassigned
in a way that has no apparent relationship to the structure of the analyzed
signal.     

The reassignment operations can be used
to gauge the quality or reliability of spectral analysis data. 
Large time or frequency reassignments indicate energy concentrated
far from the geometrical center of the analysis window. Since
window functions used in spectral analysis emphasize signal
energy near their geometrical centers and de-emphasize
signal energy far from their centers, large reassignments
indicate data derived primarily from signal features that
are not well-represented in a particular analysis window. 
Reassigned spectral data judged to be unreliable on the basis
of large time reassignments~\cite{fitz01on} or large frequency
reassignments~\cite{gardner05instantaneous} can be pruned
from the representation to further improve its readability,
with the assurance that, owing to the high redundancy 
of \stft\ data, the unreliable data will be represented
more reliably in a neighboring \stft\ frame or channel.
Gardner~\cite{gardner05instantaneous} further showed that
consensus among neighboring estimates of
reassigned frequency indicates that those frequency 
estimates are reliable, and therefore consensus can 
be used to guide the choice of an optimal analysis
window for resolving sinusoidal components in spectrally 
sparse sounds.

When there is a high degree of consensus among reassigned
frequency estimates, then the reassigned frequency, 
$\hat{\omega}(t, \omega)$ 
is changing very little with respect to the center frequency of 
the analysis window, that is
\begin{equation}
\frac{\partial \hat{\omega}(t, \omega)}{\partial \omega} \approx 0.
\end{equation}
Since, from Equation~\ref{eqn:rafreqop}, the reassigned frequency 
is computed from the partial derivative of spectral phase
with respect to time, consensus among reassigned frequency
estimates can be evaluated locally from the \emph{mixed}
partial derivative of spectral phase with respect to time and 
frequency. Specifically, in the vicinity of quasi-sinusoidal 
components, all frequencies, $\omega$, should be mapped
to approximately the same reassigned frequency, 
$\hat{\omega}(t, \omega)$, so
\begin{equation}
\label{eqn:mpd0}
\frac{\partial \hat{\omega}(t, \omega)}{\partial \omega} 
	= 1 + \frac{\partial^{2} \phi(t, \omega)}{ \partial t \partial \omega} 
	\approx 0.
\end{equation}

\begin{figure}
   \centering
   \includegraphics[width=\picwidth]{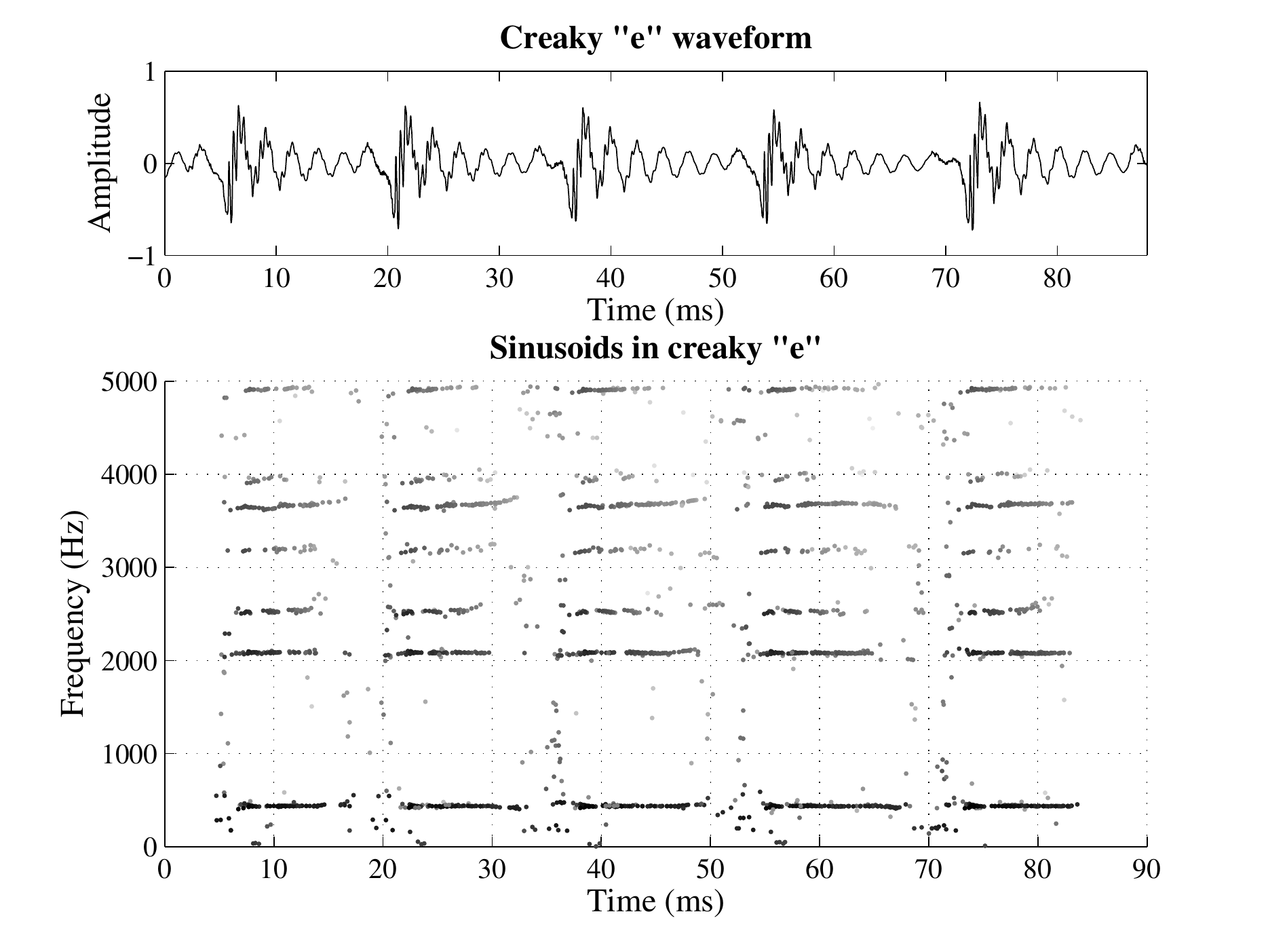} 
   \caption{Reassigned spectrogram showing only the sinusoidal components
   in the vowel \emph{e} (day).
   The spectrogram was computed using a 7.8~ms Hann window.}
   \label{fig:creakysines}
\end{figure}

Nelson showed that the mixed partial derivative of spectral phase 
can be used to clean up or ``de-speckle'' reassigned spectrograms
by removing data that does not correspond to strongly sinusoidal
or impulsive components in the analyzed 
signal~\cite{nelson02instantaneous,nelson01cross}.
By plotting just those points in a reassigned spectrogram meeting the 
condition on the mixed partial derivative expressed in Equation~\ref{eqn:mpd0},
a spectrogram showing just the strongly-sinusoidal components can be drawn.
In a speech signal, analyzed using a short analysis window, 
these will be chiefly the vocal tract resonances. A reassigned
spectrogram pruned in this way is shown in Figure~\ref{fig:creakysines}
for the same speech signal plotted in Figure~\ref{fig:creakyraw}.

\begin{figure}
   \centering
   \includegraphics[width=\picwidth]{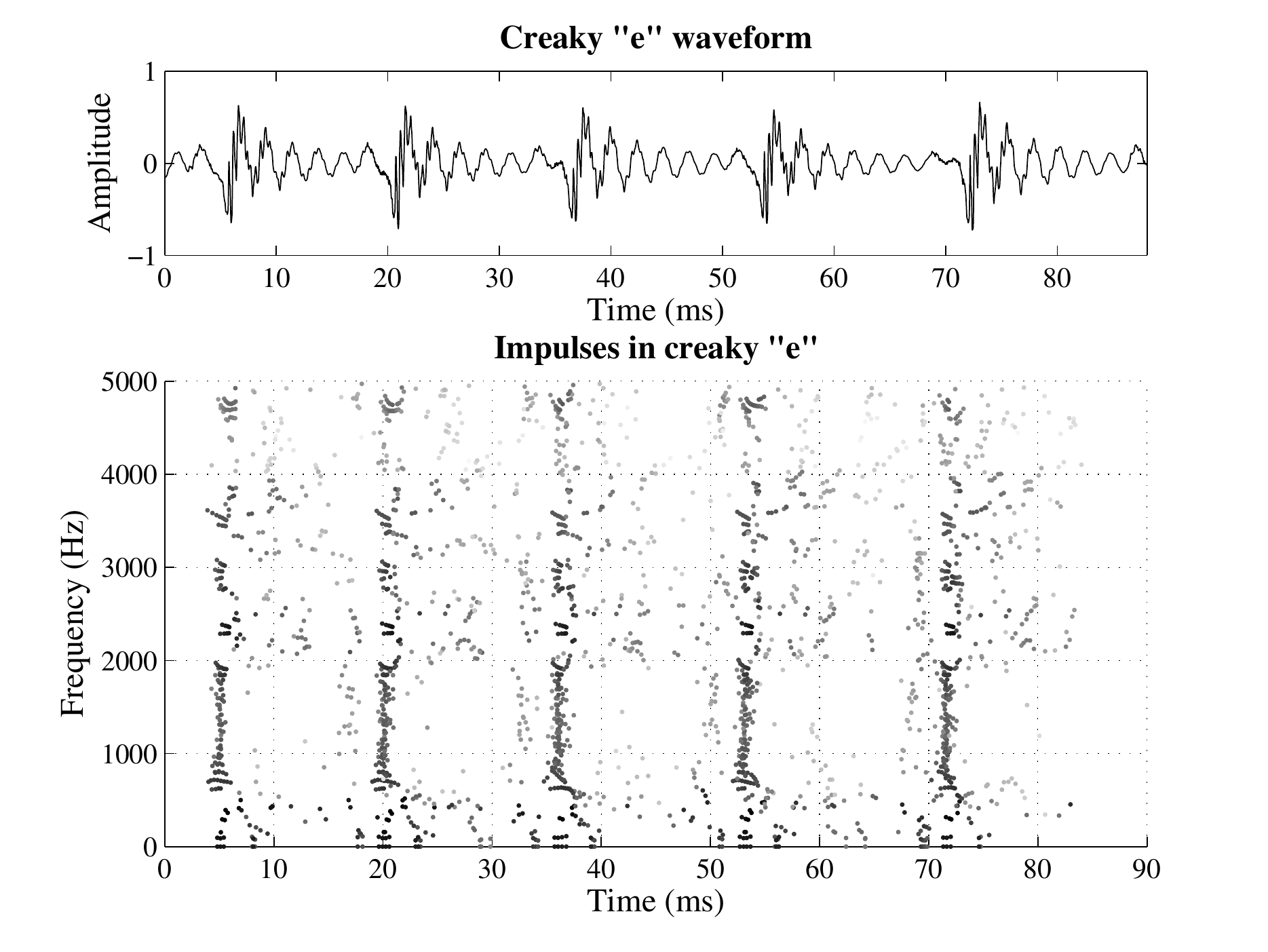} 
   \caption{Reassigned spectrogram showing only the impulsive components
   in the vowel \emph{e} (day).
   The spectrogram was computed using a 7.8~ms Hann window.}
   \label{fig:creakyimps}
\end{figure}

Nelson further demonstrated that impulsive
components in a signal should be characterized by a high
degree of consensus among neighboring reassigned time
estimates. Near the time of the impulse, the reassigned time,
$\hat{t}(t, \omega)$,
should be changing very slowly with respect to the temporal
center of the analysis window, that is,
\begin{equation}
\frac{\partial \hat{t}(t, \omega)}{\partial t} \approx 0.
\end{equation}
Since, from Equation~\ref{eqn:ratimeop}, the reassigned time 
is computed from the partial derivative of spectral phase
with respect to frequency, consensus among reassigned time
estimates can also be evaluated locally from the mixed
partial derivative of spectral phase.
In the vicinity of impulsive
components, all times, $t$, should be mapped
to approximately the same reassigned time, 
$\hat{t}(t, \omega)$, so
\begin{equation}
\label{eqn:mpd1}
\frac{\partial \hat{t}(t, \omega)}{\partial \omega} 
	= - \frac{\partial^{2} \phi(t, \omega)}{ \partial t \partial \omega} 
	\approx 0.
\end{equation}
By plotting just those points in a reassigned spectrogram meeting the 
condition on the mixed partial derivative expressed in Equation~\ref{eqn:mpd1},
a spectrogram can be drawn that clearly and precisely localizes 
the impulsive events in a signal. For example, using a short (relative 
to the fundamental period) analysis window, the individual glottal
pulses in a speech signal can be plotted, as shown for the creaky
\emph{e} signal in Figure~\ref{fig:creakyimps}.

Using the phase of the \mwt, Nelson identified the 
reassigned data corresponding to sinusoidal components
as the points satisfying
\begin{equation}
\label{eqn:mpdnelson0}
\frac{\partial^{2} \phi_{t}(\omega)}{ \partial t \partial \omega} \approx 0.
\end{equation}
and the 
reassigned data corresponding to impulsive components
as the points satisfying
\begin{equation}
\label{eqn:mpdnelson1}
\frac{\partial^{2} \phi_{t}(\omega)}{ \partial t \partial \omega} \approx 1
\end{equation}
Equations~\ref{eqn:mpdnelson0} and~\ref{eqn:mpdnelson1} 
are exactly equivalent to Equations~\ref{eqn:mpd0}
and~\ref{eqn:mpd1}, respectively; 
the apparent shift by one reflects the difference
in spectral phase reported by the \mwt\ and the \stft.

\begin{figure}
   \centering
   \includegraphics[width=\picwidth]{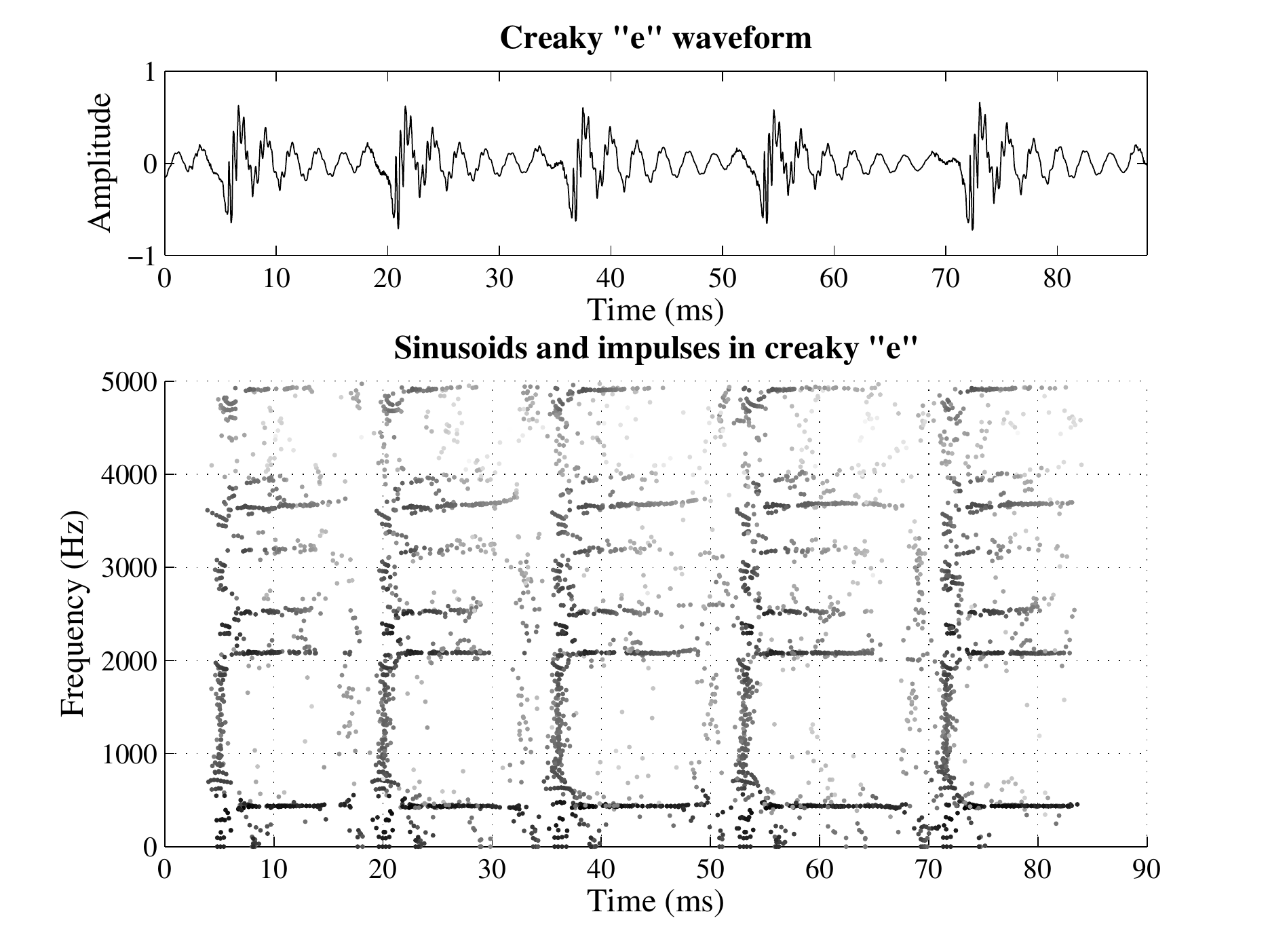} 
   \caption{``Despeckled'' reassigned spectrogram showing only the sinusoidal 
   and impulsive components in the vowel \emph{e} (day).
   The spectrogram was computed using a 7.8~ms Hann window.}
   \label{fig:creakyboth}
\end{figure}

\begin{figure}
   \centering
   \includegraphics[width=\picwidth]{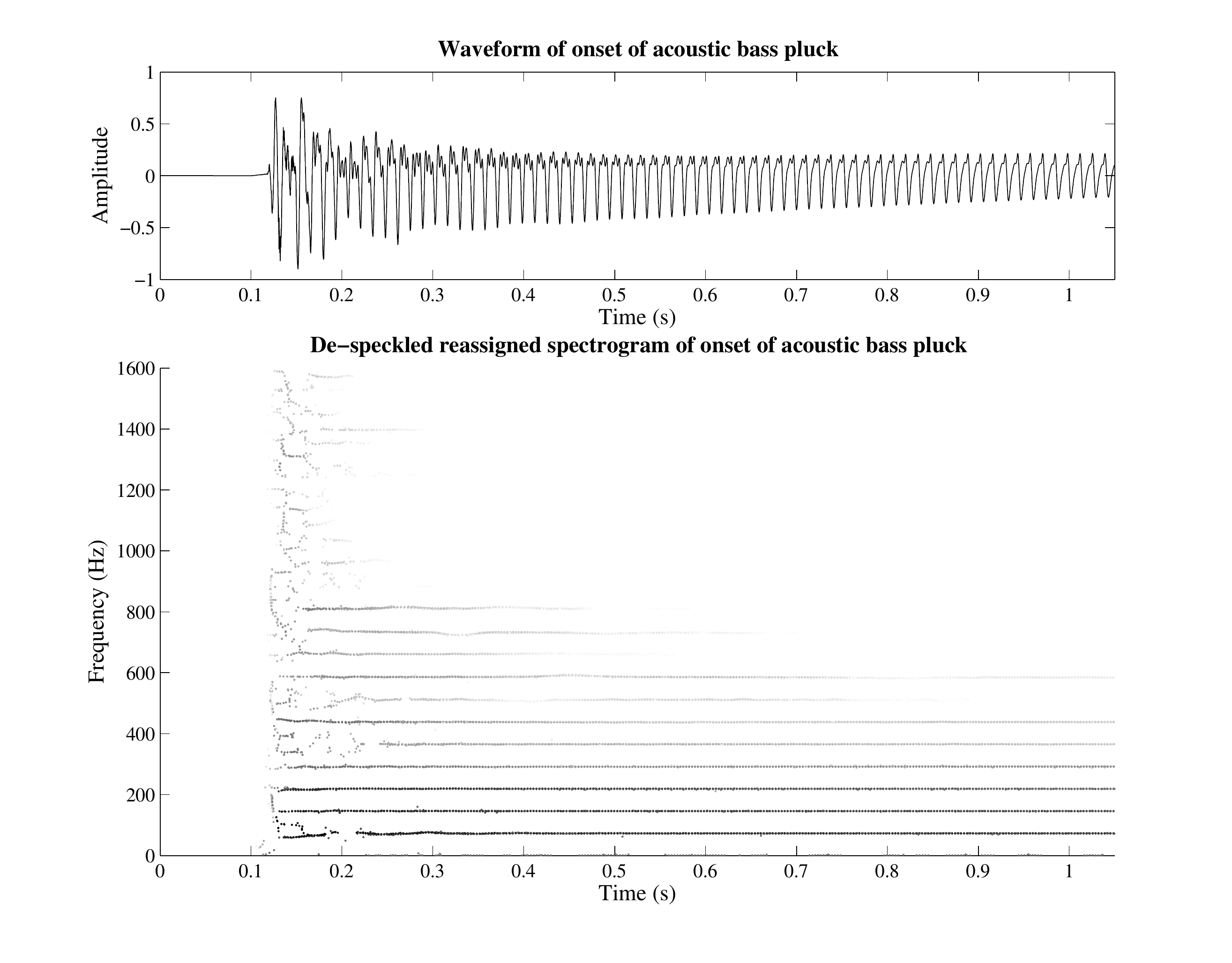} 
   \caption{``Despeckled'' reassigned spectrogram showing only the sinusoidal 
   and impulsive components in the onset 
   of an acoustic bass tone having a sharp pluck and a
   fundamental frequency of approximately 73.4~Hz. 
   The spectrogram was computed
   using a  65.7~ms Kaiser window with a shaping
   parameter of~12.    }
   \label{fig:bassdespeck}
\end{figure}

By plotting just those points in a reassigned spectrogram 
meeting the disjunction of these two conditions yields
a ``denoised'' reassigned spectrogram showing 
precisely-localized quasi-sinusoidal and impulsive components,
and excluding the objectionable ``speckle''.
Figure~\ref{fig:creakyboth} shows a fully ``despeckled''
reassigned spectrogram of the creaky \emph{e}, constructed
solely from reassigned spectral data satisfying one of the 
conditions in Equations~\ref{eqn:mpd0} and~\ref{eqn:mpd1}.
Figure~\ref{fig:bassdespeck} shows a similarly ``despeckled''
reassigned spectrogram for the acoustic bass pluck shown in 
Figure~\ref{fig:rapluck}.

Nelson used finite differences to compute
the mixed partial derivative of spectral phase, but using the
derivations of the reassignment operations in Sections~\ref{sec:derivetime}
and~\ref{sec:derivefreq}, it can be shown the the mixed partial
derivative can be computed directly from Fourier transforms by
\begin{align}
\frac{\partial ^{2} \phi(t,\omega)}{\partial t \partial \omega} 
 &= \Re \biggl \{ 
       \frac{X_{\mathcal{TD}h}(t,\omega) X^\ast(t,\omega)}{| X(t,\omega) |^{2}} \biggr \}
       - \Re \biggl \{ \frac{X_{\mathcal{T}h}(t,\omega) X_{\mathcal{D}h}(t,\omega) }{ X^{2}(t,\omega) } \biggr \} 
\end{align}
where $X_{\mathcal{TD}h}(t,\omega)$ is the \stft\ of $x(t)$ computed 
using a window $h_{\mathcal{TD}}(t) = t \frac{d}{dt}h(t)$, that is,
the window used to compute $X_{\mathcal{D}h}(t,\omega)$
multiplied by a time ramp.

\section{Phase-Correct Additive Sound Modeling}
\label{sec:phase}

In many applications, only the reassigned energy distribution is desired. 
In sound modeling applications, where the goal is to construct a model
of the sound that can be used to reconstruct the sound, possibly with
modifications, we retain not the squared magnitude of the Fourier transform,
the spectrogram, but the magnitude and phase of the transform.

\begin{figure}
   \centering
   \includegraphics[width=\picwidth]{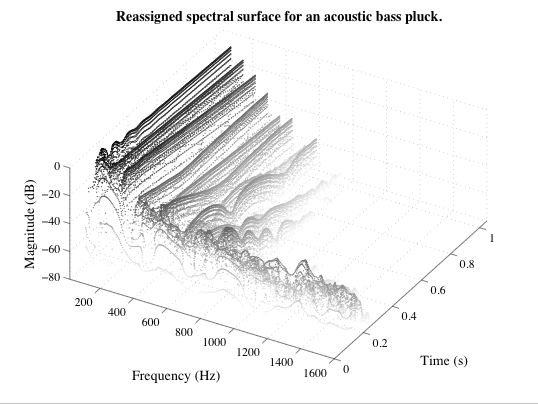} 
   \caption{Reassigned spectral surface for the onset 
   of an acoustic bass tone having a sharp pluck and a
   fundamental frequency of approximately 73.4~Hz. 
   The spectrogram was computed
   using a  65.7~ms Kaiser window with a shaping
   parameter of~12. 
   }
   \label{fig:surface}
\end{figure}

The \emph{\rabwe\ additive sound model}~\cite{fitz02cellutes} 
is a high-fidelity representation that
allows manipulations and transformations to be applied to a great
variety of sounds, including noisy and non-harmonic sounds.
It is similar in spirit to
traditional sinusoidal models~\cite{mcaulay86speech,serra90spectral,fitz96sinusoidal} in
that a waveform is modeled as a collection of components, called {\em
partials}, having time-varying frequencies and amplitudes. 
Estimates of partial frequency, amplitude, and phase
are obtained by following ridges on a reassigned
time-frequency surface, such as the one shown in 
Figure~\ref{fig:surface}, constructed by reassigning discrete \stft\ data.
This algorithm shares with traditional
sinusoidal methods the notion of temporally-connected partial
parameter estimates, but by contrast, the reassigned 
estimates are non-uniformly
distributed in both time and frequency. 
The  \rabwe\ model yields greater
resolution in time and frequency than is possible using conventional
additive techniques, and can preserve the temporal envelope of transient
signals, even in modified reconstruction, if the short-time phase information 
is properly maintained~\cite{fitz01on}. 
Preserving phase is important for reproducing 
transients and short-duration complex sounds having
significant information in the temporal 
envelope~\cite{quatieri93timescale}.

The phase reported by the \stft\ is referenced to the geometrical center of the
analysis window in time and frequency. For a sinusoid 
having instantaneous frequency equal to $\omega_{i}(t)$ (which is assumed to
be slowly-varying with respect to time), the argument of the \stft\ evaluated 
at $t,\omega_{i} (t)$ will be precisely the phase of the sinusoid at time $t$. 
If the \stft\ is evaluated at some nearby frequency, $\omega_{i} (t) + \epsilon$,
then the argument will not be precisely the phase of the sinusoid, because the
transform is equivalent to the output of a bank of linear phase
bandpass filters. These filters have phase equal to zero at the center of
their passbands, so a sinusoid having frequency equal to the center frequency 
of the filter will see no phase shift, but a sinusoid having frequency not
equal to the center frequency of the filter (but still in the passband) will 
experience a linear phase shift.

Frequency reassignment computes frequencies for discrete \stft\ data
that are equal to the instantaneous frequency of the dominant
component in the signal under analysis at each time and frequency 
at which the transform is evaluated, and attributes the data to these
reassigned frequencies. Since the reassigned data 
represent energy in signal components having frequencies that
are not at the geometrical center of the analysis window, it follows that
the data is perturbed by a linear phase shift. This perturbation is easy
to correct, because the slope of the phase response of the 
transform filters is known. In fact, since the phase is linear over the
entire passband of the filter (this is why the finite difference approximation
to the derivative is so accurate at points of significant energy, 
see section~\ref{sec:approx}), 
the phase of reassigned short-time data can be corrected 
to agree with the reassigned frequency by linear interpolation of
the discrete short-time phase spectrum.

Similarly, data that is reassigned in time away from the geometrical center of
the analysis window needs to be corrected for the phase travel due to 
sinusoidal oscillation over the interval of time reassignment (that is, the 
interval between the reassigned time and the  temporal center of the analysis 
window). In order to account for this phase travel precisely, 
the frequency trajectory must be known precisely. 
For many sounds, provided that the analysis window is not too long, 
the frequency can be assumed to be constant over the time reassignment interval,
so a shift by $\hat{\omega}(t,\omega) \cdot \left[ \hat{t}(t,\omega)-t \right]$ 
is sufficient to correct the phase of reassigned short-time data 
to agree with the reassigned time
($\hat{\omega}(t,\omega)$ is the reassigned instantaneous
frequency and $\hat{t}(t,\omega)-t$ is time reassignment interval).



\section{Computation of Higher-Order Phase Derivatives}
\label{sec:higher}

The method of reassignment uses the partial derivatives of
spectral phase with respect to time and frequency. Applications
of higher-order partial derivatives of spectral phase have
been proposed as well. Nelson showed that higher-order
partial derivatives of phase could be approximated using
cross-spectral surfaces~\cite{nelson02instantaneous,nelson01cross},
but did not discuss any applications beyond the estimation of
frequency slope, or chirp rate.
Rihaczek~\cite{rihaczek68signal} proposed that the second derivatives of the
spectral phase be used to estimate the optimal dimensions of 
the time-frequency cells in a Gabor decomposition.

By methods similar to those described in 
Sections~\ref{sec:derivetime} and~\ref{sec:derivefreq}, it
can be shown that the second partial derivative of phase
with respect to frequency and time can be computed
\begin{align}
\frac{\partial ^{2} \phi(t,\omega)}{\partial \omega^{2}} 
 &= \Im \biggl \{ \biggl ( 
       \frac{X_{\mathcal{T}h}(t,\omega) X^\ast(t,\omega)}{| X(t,\omega) |^{2}}\biggr )^{2} \biggr \}
       - \Im \biggl \{ \frac{X_{\mathcal{T}^{2}h}(t,\omega) X^\ast(t,\omega) }{ | X(t,\omega) |^{2} } \biggr \} \\
\frac{\partial ^{2} \phi(t,\omega)}{\partial t^{2}} 
&= \Im \biggl \{ \frac{X_{\mathcal{D}^{2}h}(t,\omega) X^\ast(t,\omega) }{ | X(t,\omega) |^{2} } \biggr \} 
      - \Im \biggl \{ \biggl ( 
      \frac{X_{\mathcal{D}h}(t,\omega) X^\ast(t,\omega)}{| X(t,\omega) |^{2}}\biggr )^{2} \biggr \}
\end{align}
where $X_{\mathcal{T}^{2}h}(t,\omega)$ is the \stft\ of $x(t)$ computed 
using a window $h_{\mathcal{T}^{2}}(t) = t^{2}h(t)$
and $X_{\mathcal{D}^{2}h}(t,\omega)$ is the \stft\ of $x(t)$ computed 
using a window $h_{\mathcal{D}^{2}}(t)= \frac{d^{2}}{dt^{2}}h(t)$. 

We think that higher-order phase derivatives 
might be useful in computing local estimates
of time and frequency spread that will allow us to construct 
more robust models of noisy sounds. 

\section{Conclusion}
\label{sec:conclusion}

We have presented the theory of time-frequency reassignment in the
context of the spectrogram, the most commonly-used time-frequency
representation in speech and audio processing. 
Time-frequency reassignment sharpens blurry time-frequency
data by relocating the data according to local estimates of instantaneous
frequency and group delay. This mapping to reassigned time-frequency
coordinates is very precise for signals that are 
separable in time and frequency with respect to the analysis window. 
We have discussed methods for computing
reassigned times and frequencies in digital systems, and 
offered a derivation of one popular and efficient method. We 
believe that many speech and audio processing applications
employing short-time spectral analysis could benefit from the 
straightforward application of the method of reassignment, and 
have discussed some examples from our own research. 
We further believe that extensions to the method of reassignment
to efficiently compute mixed and higher-order spectral derivatives 
may provide a means of computing other features of interest in 
speech and audio signal processing. 

\begin{ack}
The authors would like to acknowledge the
contributions of Mike O'Donnell and Lippold Haken,
whose advice and comments have greatly 
improved the quality of this manuscript.
\end{ack}


\bibliographystyle{elsart-num}
\bibliography{reassignment}

\begin{thebibliography}{10}
\expandafter\ifx\csname url\endcsname\relax
  \def\url#1{\texttt{#1}}\fi
\expandafter\ifx\csname urlprefix\endcsname\relax\def\urlprefix{URL }\fi

\bibitem{flandrin03time}
P.~Flandrin, F.~Auger, E.~Chassande-Mottin, Time-frequency reassignment: From
  principles to algorithms, in: A.~Papandreou-Suppappola (Ed.), Applications in
  Time-Frequency Signal Processing, CRC Press, 2003, Ch.~5, pp. 179 -- 203.

\bibitem{fulop05algorithms}
S.~Fulop, K.~Fitz, Algorithms for computing the time-corrected instantaneous
  frequency (reassigned) spectrogram, with applications, Journal of the
  Acoustical Society of America (2005) to appear.

\bibitem{auger95improving}
F.~Auger, P.~Flandrin, Improving the readability of time-frequency and
  time-scale representations by the reassignment method, IEEE Transactions on
  Signal Processing 43~(5) (1995) 1068 -- 1089.

\bibitem{kodera78analysis}
K.~Kodera, R.~Gendrin, C.~de~Villedary, Analysis of time-varying signals with
  small {$BT$} values, IEEE Transactions on Acoustics, Speech and Signal
  Processing ASSP-26~(1) (1978) 64 -- 76.

\bibitem{dolson86pv}
M.~Dolson, The phase vocoder: A tutorial, Computer Music Journal 10~(4) (1986)
  14 -- 27.

\bibitem{mcaulay86speech}
R.~J. McAulay, T.~F. Quatieri, Speech analysis/synthesis based on a sinusoidal
  representation, IEEE Transactions on Acoustics, Speech, and Signal Processing
  ASSP-34~(4) (1986) 744 -- 754.

\bibitem{cohen95time}
L.~Cohen, Time-Frequency Analysis, Prentice-Hall, Englewood Cliffs, NJ, 1995.

\bibitem{papoulis68systems}
A.~Papoulis, Systems and Transforms with Applications to Optics, McGraw-Hill,
  New York, New York, 1968, Ch. 7.3, p. 234.

\bibitem{rihaczek68signal}
A.~W. Rihaczek, Signal energy distribution in time and frequency, IEEE
  Transactions on Information Theory 4~(3) (1968) 369 -- 374.

\bibitem{nelson02instantaneous}
D.~J. Nelson, Instantaneous higher order phase derivatives, Digital Signal
  Processing 12~(2 -- 3) (2002) 416 -- 428.

\bibitem{nelson01cross}
D.~J. Nelson, Cross-spectral methods for processing speech, Journal of the
  Acoustical Society of America 110~(5) (2001) 2575 -- 2592.

\bibitem{flanagan66phase}
J.~L. Flanagan, R.~M. Golden, Phase vocoder, Bell System Technical Journal
  (1966) 1493 -- 1509.

\bibitem{nakatani04robust}
T.~Nakatani, T.~Irino, Robust and accurate fundamental frequency estimation
  based on dominant harmonic components, The Journal of the Acoustical Society
  of America 116~(6) (2004) 3690--3700.

\bibitem{goto00robust}
M.~Goto, A robust predominant-{F}0 estimation method for real-time detection of
  melody and bass lines in cd recordings, in: Proceedings of the IEEE
  International Conference on Acoustics, Speech, and Signal Processing, 2000,
  pp. II757 -- II760.

\bibitem{gardner05instantaneous}
T.~J. Gardner, M.~O. Magnasco, Instantaneous frequency decomposition: An
  application to spectrally sparse sounds with fast frequency modulations, The
  Journal of the Acoustical Society of America 117~(5) (2005) 2896--2903.

\bibitem{plante98improvement}
F.~Plante, G.~Meyer, W.~A. Ainsworth, Improvement of speech spectrogram
  accuracy by the method of spectral reassignment, IEEE Transactions on Speech
  and Audio Processing 6~(3) (1998) 282 -- 287.

\bibitem{hainsworth01analysis}
S.~Hainsworth, M.~Macleod, P.~Wolfe, Analysis of reassigned spectrograms for
  musical transcription, in: IEEE Workshop on the Applications of Signal
  Processing to Audio and Acoustics, 2001, pp. 23 -- 26.

\bibitem{fitz01on}
K.~Fitz, L.~Haken, On the use of time-frequency reassignment in additve sound
  modeling, Journal of the Audio Engineering Society 50~(11) (2002) 879 -- 893.

\bibitem{serra90spectral}
X.~Serra, J.~O. Smith, Spectral modeling synthesis: A sound analysis/synthesis
  system based on a deterministic plus stochastic decomposition, Computer Music
  Journal 14~(4) (1990) 12 -- 24.

\bibitem{keiler02survey}
F.~Keiler, S.~Marchand, Survey on extraction of sinusoids in stationary sounds,
  in: Proc. DAFx-02 Digital Audio Effects Conference, Hamburg, 2002, pp.
  51--58.

\bibitem{auger94why}
F.~Auger, P.~Flandrin, The why and how of time-frequency reassignment, in:
  Proceedings of the IEEE-SP International Symposium on Time-Frequency and
  Time-Scale Analysis, 1994, pp. 197 -- 200.

\bibitem{charpentier86pitch}
F.~Charpentier, Pitch detection using the short-term phase spectrum, in: IEEE
  International Conference on Acoustics, Speech, and Signal Processing,
  Vol.~11, 1986, pp. 113 -- 116.

\bibitem{abe96robust}
T.~Abe, T.~Kobayashi, S.~Imai, Robust pitch estimation with harmonics
  enhancement in noisy environments based on instantaneous frequency, in:
  Proceedings of the Fourth International Conference on Spoken Language,
  Vol.~2, 1996, pp. 1277 -- 1280.

\bibitem{fitz02cellutes}
K.~Fitz, L.~Haken, S.~Lefvert, C.~Champion, M.~O'Donnell, Cell-utes and
  flutter-tongued cats: Sound morphing using {L}oris and the reassigned
  bandwidth-enhanced model, Computer Music Journal 27~(4) (2003) 47 -- 65.

\bibitem{fitz96sinusoidal}
K.~Fitz, L.~Haken, Sinusoidal modeling and manipulation using {L}emur, Computer
  Music Journal 20~(4) (1996) 44 -- 59.

\bibitem{quatieri93timescale}
T.~F. Quatieri, R.~B. Dunn, T.~E. Hanna, Time-scale modification of complex
  acoustic signals, in: Proceedings of the International Conference on
  Acoustics, Speech, and Signal Processing, IEEE, 1993, pp. I--213 -- I--216.

\end{thebibliography}






\textbf{Kelly Fitz} received a Ph.D. in Electrical Engineering from the 
University of Illinois at Urbana-Champaign in 1999. He was the principle
developer of the sound analysis and synthesis software \emph{Lemur},
and, at the National Center for Supercomputing Applications, 
co-developer of the \emph{Vanilla Sound Server}, an application providing real-time
sound synthesis for virtual environments. Dr. Fitz is currently 
Assistant Professor of Electrical Engineering and Computer Science
at Washington State University, and the principle
developer of the \emph{Loris} software for sound modeling and morphing. 
His research interests include speech and audio processing, 
digital sound synthesis, and computer music composition.

\textbf{Sean A. Fulop} was born in Calgary, Alberta, on 16 March 1970.
He received a Ph.D. in linguistics from UCLA in 1999.  
After holding lectureships at San Jose State, the University of Chicago, 
and California State University Fresno, he became an Assistant Professor 
at Fresno in 2005.  
Dr. Fulop's research interests include applications of new DSP methods in 
the analysis of speech and animal communication, as well as mathematical 
modeling of human language with the aid of computational learning theory and logic. 
Nonacademic interests include progressive rock music and nurturing an old sports car.

 \end{document}